\documentclass[AMA,STIX1COL]{WileyNJD-v2}

\articletype{Research Article}%

\makeatletter  
\newif\if@restonecol

\makeatother  
\usepackage[linesnumbered,ruled,vlined]{algorithm2e}
\usepackage{caption}
\usepackage{subcaption}
\usepackage{graphicx}
\usepackage{amsmath}
\usepackage{amssymb}
\usepackage{textcomp}
\usepackage{booktabs}
\usepackage[justification=justified]{caption}
\usepackage{color} 
\usepackage{float}
\usepackage{hyperref} 
\usepackage{url}
\usepackage{multirow}
\usepackage{makecell}

\articletype{Article Type}%

\received{<day> <Month>, <year>}
\revised{<day> <Month>, <year>}
\accepted{<day> <Month>, <year>}

\begin{document}

\title{\textcolor{black}{LSRAM: A Lightweight Autoscaling and SLO Resource Allocation Framework for Microservices Based on Gradient Descent}}

\author[1]{Kan Hu}

\author[1]{Minxian Xu}

\author[1]{Kejiang Ye}

\author[2]{Chengzhong Xu}

\authormark{K. HU, M. XU \textsc{et al}}

\address[1]{\orgdiv{Shenzhen Institute of Advanced Technology}, \orgname{Chinese Academy of Sciences}, \orgaddress{\state{Shenzhen}, \country{China}}}
\address[2]{\orgdiv{State Key Lab of IOTSC, Department of Computer Science}, \orgname{University of Macau}, \orgaddress{\state{Macau SAR}, \country{China}}}



\corres{Minxian Xu, Shenzhen Institute of Advanced Technology, Chinese Academy of Sciences, Shenzhen, China. \email{mx.xu@siat.ac.cn}}



\abstract[Abstract]{Microservices architecture has become the dominant architecture in cloud computing paradigm with its advantages of facilitating development, deployment, modularity and scalability. The workflow of microservices architecture is transparent to the users, who are concerned with the quality of service (QoS). Taking Service Level Objective (SLO) as an important indicator of system resource scaling can effectively ensure user's QoS, but how to quickly allocate end-to-end SLOs to each microservice in a complete service so that it can obtain the optimal SLO resource allocation scheme is still a challenging problem. Existing microservice autoscaling frameworks based on SLO resources often have heavy and complex models that demand substantial time and computational resources to get a suitable resource allocation scheme. Moreover, when the system environment or microservice application changes, these methods require significant time and resources for model retraining. In this paper, we propose LSRAM, a lightweight SLO resource allocation management  framework based on the gradient descent method to overcome the limitation of existing methods in terms of heavy model, time-consuming, poor scalability, and difficulty in retraining. LSRAM has two stages: at stage one, the lightweight SLO resource allocation model from LSRAM can quickly compute the appropriate SLO resources for each microservice; \textcolor{black}{at} stage two, LSRAM's SLO resource update model enables the entire framework to quickly adapt to changes in the cluster environment (e.g. load and applications). Additionally, LSRAM can effectively handle bursty traffic and highly fluctuating load application scenarios. Compared to state-of-the-art SLO allocation frameworks, LSRAM not only guarantees users' QoS but also reduces resource usage by 17\%.}


\maketitle


\section{Introduction} \label{introduction}

In contrast to the traditional monolithic architecture, where all functional components are integrated, microservices architecture decouples a complex application into multiple independent and lightweight distributed components \cite{zeng2023topology}. Each component is responsible for specific functionalities, facilitating efficient management, maintenance, and updates \cite{gan2019open}. Therefore, the use of microservices architecture in cloud computing paradigm has become a trend, widely adopted by major enterprises such as Amazon \cite{amazon2015} Netflix \cite{netflixtechblog2021}, Alibaba \cite{luo2021characterizing} \cite{xu2024practice}. Due to the lightweight and loosely coupled nature of microservices architecture, each functional component operates independently \cite{luo2022erms}. As the architecture handles increasing workloads, system administrators can identify the components bearing the load and independently scale them, rather than expand resources for the entire cloud service application. 

For users of cloud service applications, their primary concern lies not in the specific operational details of backend services but in their own service experience. Users have varying tolerance levels for response time (RT) across different applications. For instance, Amazon found every 100ms of latency would cost them 1\% in sales, and Google conducted a study revealing that a delay of 0.5 seconds in the RT of a search page could lead to a 20\% decrease in traffic~\cite{amazon2023}. Consequently, cloud service providers establish SLO metric \cite{delimitrou2018amdahl}, imposing constraints on the mean or tail latency distribution of end-to-end latency within cloud services. This is achieved to ensure QoS \cite{jawaddi2022review} remains within acceptable ranges.

\textcolor{black}{Cloud service users prioritize their QoS, leading to numerous studies focusing on the allocation of SLOs \cite{mirhosseini2019queuing}. These studies define a partial SLO that represent expected targets for latency in specific parts among the application for individual microservice components to ensure the end-to-end SLO is met if all partial SLOs are satisfied \cite{urgaonkar2008agile}. Existing methodologies involve the definition of partial SLOs based on empirical knowledge, reinforcement learning techniques, or the application of deep learning algorithms to compute and design partial SLOs \cite{khorsand2018fahp}. However, a significant limitation of these methods is that neural networks tend to become excessively large with the increase in the microservices topology. In addition, a single microservice can be shared among multiple services, each with diverse workload patterns and different SLO, for such shared microservices, it is challenging to define their SLO resource allocation. These result in slow computation speeds and substantial consumption of system resources. Additionally, these approaches struggle to effectively respond to sudden surges in traffic.}

To address the above limitations, we propose LSRAM, a \underline{L}ightwight \underline{S}LO \underline{R}esource \underline{A}llocation \underline{M}anagement framework for highly volatile loads environments based on SLO allocation. LSRAM takes the load latency characteristics of microservices and the relationship between microservices as model inputs, rapidly derive allocation scenarios through the lightweight SLO resource allocation model. During runtime, LSRAM continuously monitors changes in microservice applications and loads in the cluster, updating the SLO resources of each microservice in real-time to make them more suitable for the current cluster environment. For highly volatile loads, LSRAM implements a prediction model which can predict next time period load in advance and respond quickly to resource scaling, while further considering cascading dependencies and taking instance creation time into account in the resource scaling architecture to further ensure the QoS for users. 

\textcolor{black}{Compared to traditional methods of CPU and memory resource allocation, treating SLO resources as a system resource allocated to each microservice more effectively ensures the quality of user services. While an appropriate CPU utilization threshold can improve overall system resource efficiency, it cannot effectively guarantee the quality of user services. The metrics and optimization goals of these two approaches diverge from the outset: CPU resource allocation aims to maintain a high level of system resource utilization, whereas SLO resource allocation focuses on maximizing system resource savings while ensuring user service quality. Moreover, through an effective SLO resource allocation model, we can identify the optimal SLO resource allocation scheme for the current environment, ensuring that microservices minimize system costs while maintaining service quality.}

In addition, we built a prototype system of LSRAM based on Kubernetes, and evaluated LSRAM by deploying realistic microservice applications and conducting high fluctuation load test evaluations, and the experimental results show that LSRAM can reduce the CPU resource consumption by about 17\% compared with the existing methods, and has more significant performance in reducing SLO violations. Our key contributions are as follows:

\begin{itemize}
    \item [$ \bullet $] We design a lightweight tail-latency based SLO resource allocation model, and proposed appropriate computational methods for shared microservices. This model can quickly compute a SLO resource allocation scheme. 
    \item [$ \bullet $] We propose a SLO resource update model that allows the SLO resource allocation scheme to be updated in real-time based on load traffic changes in the cluster during runtime, ensuring the effectiveness of the system's runtime;
    \item [$ \bullet $] We optimize a neural network algorithm for load prediction, enhancing the framework's ability to cope with bursty traffic. This optimization effectively guarantees end-to-end SLOs of microservices and improves user experience. Additionally, we have designed a high volatility mode to effectively cope with high volatility loads to ensure QoS.
\end{itemize}

The remainder of this paper is organized as follows: Section~\ref{related work} discusses the related work, which focuses on the related research work on autoscaling techniques for SLO resource management in microservices architecture. Section~\ref{approach overall} presents the general framework of LSRAM. In Section~\ref{models}, the SLO resource allocation model is described in detail, and the entire LSRAM workflow is presented. Section~\ref{experimental} introduces the settings of the experiment and the analysis of the results. Section~\ref{conclu} summarizes the main findings of the paper and outlines future work plans.

\section{RELATED WORK} \label{related work}

Autoscaling \cite{hassan2022systematic} techniques for resource management in microservices architecture is a popular and attractive topic. Due to the importance of SLO metrics in cloud services, the study of SLO-oriented and guaranteeing QoS for users plays an important role in microservice autoscaling techniques. In the following subsections, we summarize various methods for ensuring SLOs across different types of scaling approaches.

\subsection{Rule-based Autoscaling Approaches to Satisfy SLOs}

Rule-based autoscaling approaches use the utilization of predefined thresholds and heuristics to initiate scaling actions \cite{vayghan2021kubernetes} \cite{srirama2020application}. Most cloud platforms, such as Google Cloud and Amazon Web Services, employ this approach. A well-known method is Horizontal Pod Autoscaling (HPA) integrated with the Kubernetes platform \cite{k8sautoscaler}, which scales resources by defining an average CPU utilization threshold for microservices. This approach is simple and straightforward, the key is to find an appropriate threshold point to balance the relationship between RT and resource consumption. Most cloud platforms configure their thresholds based on historical experience. Google's autoscaling framework Autopilot \cite{rzadca2020autopilot}, which monitor the RT of the request to access the cloud service in real-time, when the RT is much larger or less than the predefined SLO threshold, the framework will automatically change the number of instances in a timely manner to expand or decrease the number of services, and meet the SLO indicators with the lowest resource consumption. 

To better adapt the scheduling threshold to the cluster load environment, many studies propose a dynamic threshold approach, which scales microservices with different thresholds under varying load conditions. Liu et al. \cite{liu2018fuzzy} propose a multi-model controller that inherits adaptive decisions from multiple models to select the best scenario, the main advantage of the empirical model is that it obtains high quality performance predictions based on empirical data. For new application scenarios, this controller takes other models or heuristics as a starting point and continuously optimises this empirical value by continuously incorporating all performance data into the knowledge base of the empirical model.

The main goal of these SLO-oriented methods is to determine the optimal threshold value, but they face challenges in handling intricate workload dynamics, adapting to rapidly changing conditions, and optimizing resource allocation under complex scenarios.

\subsection{Learning-based Autoscaling Approach to Satisfy SLOs}

There exists a large number of learning-based methods for autoscaling to satisfy SLOs. We summarise these methods into two categories based on the objects they learn from.

\textbf{Historical data driven learning-based methods}: These approaches focus on analysing and modelling historical data in a microservices architecture \cite{song2022automatic} \cite{li2021joint}. 
They devise a series of centralized, data-driven models to address resource scaling for microservices. Seer \cite{gan2019seer} analyzes massive amount of tracing data collected by cloud systems to learn spatial and temporal patterns, which are then translated into QoS violations using machine learning. Sage \cite{gan2021sage} is a supervised machine learning-based system that models the dependencies between the microservices using causal Bayesian networks to identify the root cause of SLO violations. While Sage can mitigate SLO violations for microservices, it lacks the ability to adapt to changes in workloads. However, Sage can predict potential QoS violations based on the underlying data it monitors and allocate resources in advance to ensure QoS. 

\textbf{Environmental interaction exploration driven learning-based methods}: \textcolor{black}{This category of research focuses on autonomously learning optimal scaling strategies by iteratively exploring interactions with the environment \cite{yu2022dependable} \cite{chen2022resource} \cite{jian2023drs}. Coscal \cite{xu2022coscal} takes advantage of data-driven decision making by efficiently learning the scaling techniques for system resources and improving the RT of scaling techniques. However, Coscal relies heavily on Q-learning, the speed and accuracy of the results it generates decrease dramatically as the amount of data increases. AutoMan\cite{cai2023automan} introduces a resource allocation strategy employing multi-agent deep deterministic policy gradient reinforcement learning. This method aims to ensure that microservices meet end-to-end tail latency SLOs. FIRM \cite{qiu2020firm} is an intelligent fine-grained resource management framework, focuses on finding and removing contention in the microservice that is the critical cause of latency SLO violation. However, FIRM lacks the ability to identify configurations, which may result in excessive resource allocation to such microservices, leading to resource wastage. GRAF \cite{park2021graf} proposes a graph neural network-based proactive resource allocation framework that minimizes overall CPU resources in the microservice chain while satisfying latency SLO. However, when faced with large-scale microservice architectures, GRAF can be inefficient and unable to achieve real-time allocation, while its inability to resolve situations related to microservice resource contention. In addition, its resource allocation scheme is not an optimal solution.}

In summary, learning-based autoscaling approaches to ensure SLO mostly share some common drawbacks including the inability to quickly adapt to changes in microservice application architectures and poor scalability. Many models consume a lot of resources for computation, which is not efficient for highly fluctuating environment.

\begin{table}[ht]
\centering
\caption{Comparison of Related Work.}
\label{tab:related work}
\resizebox{\linewidth}{!}{%
\begin{tabular}{ccccccc}
\hline
\multirow{2}{*}{\textbf{Work}} & \multirow{2}{*}{\textbf{\makecell{Load Prediction}}} & \multicolumn{2}{c}{\textbf{Model Complexity}} & \multirow{2}{*}{\textbf{\makecell{Updated Resource \\ Allocation}}} & \multirow{2}{*}{\textbf{\makecell{SLO \\ Guarantee}}} & \multirow{2}{*}{\textbf{\makecell{SLO as \\ Allocation Resources}}} \\
\cmidrule(r){3-4}
& & \textbf{Heavy} & \textbf{Lightweight} & & & \\
\hline
HPA \cite{qu2018auto}                &                 &                  & \checkmark  &                                 &                 & \\
FuSys \cite{liu2018fuzzy}            &                 &                  & \checkmark  & \checkmark                      &                 & \\
Autopilot \cite{rzadca2020autopilot} &                 &                  & \checkmark  &                                 &                 & \\
Seer \cite{gan2019seer}              & \checkmark      & \checkmark       &             &                                 &                 & \\
Sage \cite{gan2021sage}              & \checkmark      & \checkmark       &             &                                 & \checkmark     &  \\
Coscal \cite{xu2022coscal}           & \checkmark      & \checkmark       &             &                                 &                 & \\
FIRM \cite{qiu2020firm}              &                 & \checkmark       &             & \checkmark                      &                &  \\
AutoMan \cite{cai2023automan}        & \checkmark      & \checkmark       &             &                                 & \checkmark     &  \\
GRAF \cite{park2021graf}             & \checkmark      & \checkmark       &             &                                 & \checkmark      & \\
\textcolor{black}{Epma} \cite{fourati2022epma}          &                 &                  & \checkmark  &                                 &                &  \\
Hansel \cite{yan2021hansel}          &                 &                  & \checkmark  &                                 &                &  \\
Showar \cite{baarzi2021showar}       &                 &                  & \checkmark  &                                 &                 & \\
GrandSLAm \cite{kannan2019grandslam} &                 &                  & \checkmark  &                                 &                 &  \checkmark\\
Erms \cite{luo2022erms}              & \checkmark      & \checkmark       &             &                                 & \checkmark     &  \\
Parslo \cite{mirhosseini2021parslo}  &                 & \checkmark       &             & \checkmark                      &                & \checkmark \\
LSRAM (This paper)                   & \checkmark      &                  & \checkmark  & \checkmark                      & \checkmark    &   \checkmark \\
\hline
\end{tabular}%
}
\end{table}

\subsection{Model-based Autoscaling Approaches to Satisfy SLOs}

Some studies also employ formal mathematical models to establish relationships between workload characteristics and resource provisioning \cite{gias2019atom,zhao2020rhythm}.

\textcolor{black}{Epma \cite{fourati2022epma} designs two modules to detect and identify the root cause of the overload state at the application and a variety of scheduling strategies to address the identified problems. However, the resource allocation scheme given is not the optimal solution for the system, and the heuristic algorithm makes it unable to cope with unexpected loads.
Hansel \cite{yan2021hansel} analyzes the historical resource usage of microservice container pods and models it alongside service quality monitoring SLO information. The model is utilized for proactive scaling control and reactive scaling control decisions based on current SLOs. However, the models of Hansel need constant updates for training, which cannot quickly incorporate changes in microservice applications, and its resource allocation tends to over-scale to guarantee SLOs. }
Showar \cite{baarzi2021showar} models and analyzes microservice resource scaling in two directions: vertically using the empirical variance of historical resource usage to find the optimal size, and horizontally using cybernetic methods to find the optimal configuration.
GrandSLAm \cite{kannan2019grandslam} estimates the completion time for a single microservice component to propagate a request in a microservice application and uses this estimate to drive the scaling of system resources. However, it does not consider coordination, and its estimate may not be optimal.
Luo et al. \cite{luo2022erms} introduce Erms, a resource management system designed to ensure SLOs in shared microservice environment. Erms profiles microservice latency as a piece-wise linear function of the workload, resource usage, and interference. It constructs resource scaling models to optimally define latency targets for microservices with intricate dependencies. However, Erms tends to over-provision resources for online services with highly dynamic dependency graphs and requires explicit modeling for each graph, making it challenging to achieve comprehensiveness.

\textcolor{black}{Unlike the aforementioned work focusing on autoscaling resources to ensure SLO, our approach considers SLO as the allocation resources to guarantee the SLO of different microservices.
Our work is most similar to Parslo \cite{mirhosseini2021parslo}, a Gradient Descent-based approach to assign partial SLOs among nodes in a microservice graph under an end-to-end latency SLO. Parslo isolates different nodes in the Directed Acyclic Graph (DAG) from each other, enabling each microservice to scale independently through its autoscaling framework. It employs an approximate optimal partial SLO allocation scheme for minimizing overall deployment costs in general microservice graphs. However, for larger microservice call graph structures, the iteration time increases, requiring substantial resources for both offline and online SLO allocation verification. Parslo necessitates similar load-latency profiles for all microservices, yielding effective results for specific structures but limiting applicability in diverse scenarios. It is not suitable for high dynamic loads and lacks a robust strategy for sudden changes in traffic.}

\color{black}Table \ref{tab:related work} shows the comparison between our work and the related work. The main differences between LSRAM and these approaches lie in three aspects. Firstly, LSRAM designs a lightweight SLO resource allocation approach that can quickly obtain the resource allocation scheme. Secondly, LSRAM quickly adapts to changes in the cluster environment and microservice applications by the update algorithm. Thirdly, LSRAM enhances adaptability to dynamic loads and traffic bursts with load prediction.\color{black}

\section{SYSTEM MODEL} \label{approach overall}
In this section, we will show the design of the overall system architecture of LSRAM, as shown in Figure.~\ref{fig:overall}, the key modules are as follows: 

\begin{figure}[htbp]
\centering
\includegraphics[width=0.8\linewidth]{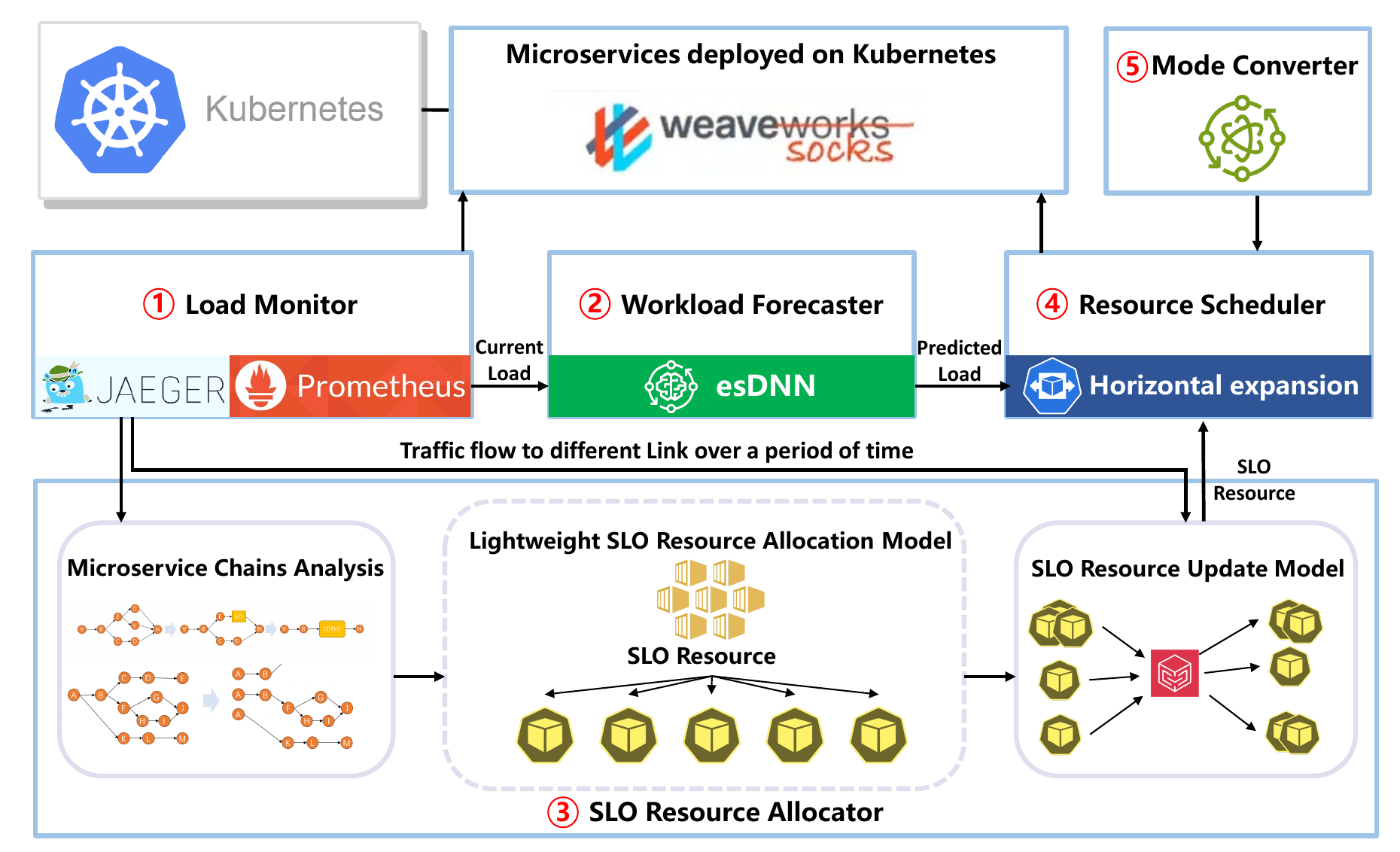}
\caption{System Model of LSRAM.}
\label{fig:overall}
\end{figure}

\textbf{Load Monitor}: The Load Monitor is designed to gather real-time load data for the current microservice deployed in cluster (via Prometheus\footnote{https://prometheus.io/}). It generates a dependency graph for each microservice (via Jaeger\footnote{https://www.jaegertracing.io/}) and tallies the number of requests flowing to various branches of the microservice over a specified time period. During the initialization phase, the load monitor senses the topology of the entire microservice and sends it to the SLO resource allocator to assist with the initial allocation of SLO resources. During runtime, the load monitor continuously collects the number of requests flowing through different microservice chains and sends the traffic distribution to the SLO resource allocator regularly. It serves as the top-level awareness of the entire framework, providing inputs necessary for quickly and accurately representing the system status, \textcolor{black}{since we only invoke APIs of the relevant tools, the load monitoring tool and the tracing tool in this component can be easily replaced with other tools offering similar functionalities.}

\begin{algorithm}[h]
    \caption{LSRAM overall scheduling algorithm.}
    \label{alg:overall}
    \KwIn{Current load $CL$}
    \KwOut{Scaling strategies to meet SLO}

\color{black}    \textbf{SLO resource initialization allocation:}\color{black}\\
    \ \ \ Calling Algorithm \ref{alg:initial} to initialize SLO allocation\;
    \ \ \ SLO resources $s_i$ allocated to each microservice\;
\color{black}    \textbf{Initialize parameters:}\color{black}\\
    \ \ Current Load $CL$, future load $FL$, time index $t$, clustering oscillation factor $O$, predefined oscillation coefficient thresholds $OCT$, predefined resource reallocation time thresholds $M_t$, number of instances $instance_i$ of each microservice, the load scaling threshold $f(s_i)$ corresponds to the SLO resources $s_i$.\\
\color{black}    \textbf{Cluster runtime:}\color{black}\\
    $O = \frac{\sqrt{\frac{1}{n}\sum_{i=1}^{n}[CL_{i}-\bar{CL}]^2}}{\bar{CL}}$ \\
    \If{$O \le OCT$}{
        Each microservice $i$ autoscaling by $s_i$\;
        $FL \leftarrow$ Predicted by esDNN model $\leftarrow CL$\\
        \If{$FL \geq f(s_i)$}{
            $instance_i = \lceil \frac{FL}{f(s_i)} \rceil$ \\
            Scaling operation\;
        }
    }
    \textbf{Update:}\\
    \If{$t \geq M_t$ or cluster environment changes}{
        Calling Algorithm \ref{alg:update} to update SLO resource\;
    }
    \textbf{Mode Converter:}\\
    \If{$O \geq OCT$ or cluster environment changes}{
        Adjusting scaling strategies to meet user service satisfaction\;
    }
    \Else{
        Keep current scaling strategies\;
    }
\end{algorithm}

\textbf{Workload Forecaster}: In order to cope with the changes of load traffic in the cluster and to enhance the system's ability for handling unexpected traffic, we design a workload Forecaster for predicting the load changes, and pre-allocating resources in advance to ensure the QoS. The Workload Forecaster optimizes our previous work, named esDNN \cite{xu2022esdnn}, a deep learning network based on supervised learning, by intensive training for highly volatile loads and bursty traffic loads. Additionally, we deploy esDNN as the prediction model to forecast the load of the next time slice by processing cluster load data obtained from the Load Monitor. \textcolor{black}{To be noted, this prediction model can be easily replaced with other lightweight prediction models.} 

\textbf{SLO Resource Allocator}: The lightweight and accurate allocation of SLO resources is one of the core functions of the LSRAM framework. It comprises microservice chain analyzer, a lightweight SLO resource allocation model, and a SLO resource update model. These components efficiently compute the allocation of SLO resources based on end-to-end tail latency (eTL), leveraging load-latency-profile (LLP, \textcolor{black}{which indicates the relationship between the request arrival rate and the eTL}) graphs and microservice chain structure diagrams. Among these, the lightweight SLO resource allocation model based on eTL is the core model of LSRAM, which solves the shortcomings of other methods in calculating the SLO allocation scheme which is time-consuming and consumes a lot of computational resources. After initialization, the existence of SLO resource update model can make LSRAM quickly adapt to the changes in the cluster environment and ensure user satisfaction, during the subsequent runtime of the framework, SLO Resource Allocator will weight each microservice according to the load traffic direction in a period of time sent by the Load Monitor and reallocate SLO resources, so as to make the whole framework system more adaptable to the current system traffic changes, ensure service quality and save system resources at the same time.

\textbf{Resource Scheduler}: Resource scheduling is based on the resource computed by SLO Resource Allocator as the scaling threshold, and the QoS is guaranteed based on the number of load-scaling instances predicted by Workload Forecaster for the next time slice. When the predicted load exceeds the allocated resource threshold, the resource scheduler scales the microservice instance ahead of time based on the number of thresholds exceeded. The Resource Scheduler will allocate the scheduling resources in advance according to the predicted value, so that when the load arrives in the next time period, there are sufficient seasonal celery resources to cope with it and ensure the QoS. 

\textbf{Mode Converter}: The Mode Converter is designed to cope with specific high-volatility load conditions. Equipped with a built-in oscillation-aware function, it monitors real-time load obtained from the Load Monitor. This function collects load values over a period of time, and when the oscillation value surpasses a predefined threshold, LSRAM identifies the cluster as being in a high-volatility state. Subsequently, the Mode Converter proactively switches the system resource scaling mode, rapidly scaling resources to meet user service demands.

LSRAM efficiently allocates suitable SLO resources to each microservice using the above mentioned components, and accurately predicts upcoming load changes to proactively allocate resources during runtime. This framework resolves the issue of excessive time and computational resource consumption associated with the eTL SLO allocation method. Instead, it introduces a lightweight computational approach that effectively handles bursty traffic and significant load fluctuations. \textcolor{black}{Moreover, each module is independent of the others, so besides the SLO resource allocator, all other modules can be easily replaced to ensure flexibility of our proposed framework}.

\section{Algorithms Developed in LSRAM} \label{models}

In this section, we introduce the overall workflow of LSRAM and focus on the algorithms deployed in LSRAM, including the lightweight SLO resource allocation model and the SLO resource update algorithm. 

\subsection{Workflow of LSRAM}

The overall workflow of LSRAM is illustrated in Algorithm~\ref{alg:overall}. LSRAM allocates appropriate SLO resources $s_i$ for each microservice based on the microservice structure provided by the load monitor through the SLO resource initial allocation algorithm (lines 1-3), laying the groundwork for resource autoscaling. During regular runtime (lines 6-15), the load monitor transmits the monitored load situation of the current cluster to the workload predictor to predict the load in the next time period. The resource scheduler pre-allocates resources based on the predicted load value. After some time periods or cluster environment changes, the SLO resource allocator executes the SLO resource update model to update the SLO resources for shared microservices and other microservices (lines 16-19). If the mode converter detects a high load fluctuation (lines 20-23), it adjusts the resource scheduling strategy to ensure user service satisfaction.

\textcolor{black}{\textbf{Complexity Analysis:} }
\textcolor{black}{Since Algorithm 1 is the scheduling algorithm for the entire system framework, its time complexity is composed of the time complexities of its various components. First, the SLO Resource Initialization Allocation has a time complexity of $O(n\log n)$ (detailed in Section~\ref{allocate}). The Cluster Runtime and mode converter have a time complexity of $O(1)$ , where they collect data over a period and compare it with the oscillation coefficient threshold, We utilized a queue to store the data. The Update phase has a time complexity of $O(m*n)$, where $m$ denotes the number of microservices that need be updated. Therefore, the total time complexity of Algorithm 1 is $O(n\log n + m*n)$.}

\subsection{Lightweight SLO Resource Allocation Model}

Microservices, owing to their loosely coupled structure, often result in user requests traversing one or more microservice chains to obtain the desired result. Each microservice within these chains operates at a distinct processing speed and may handle requests in parallel, each with its own latency. Ensuring that each microservice component consumes minimal system resources while meeting the user's RT requirements poses a significant optimization challenge. This gives rise to the following optimization problem:
\begin{equation}
min\sum_{i}^{n} C_i \ \text subject \ \text to\ \ \text{Tail}\_{\text{latency}_k} \leq SLO_k,  \label{eq: overall objective}
\end{equation}
where $C_i$ denotes the total resources consumed by each microservice, $\text{Tail}\_{\text{latency}_k}$ denotes the eTL of each chain $k$, and $SLO_k$ indicates a set RT constraint.
\textcolor{black}{The overarching optimization objective is to minimize the total amount of resources consumed by the microservices while ensuring the minimization of end-to-end latency violations. The symbols and definitions used in the paper are introduced in Table~\ref{tab:symbols}.}

\begin{table}[ht]
\centering
\caption{Symbols and Definitions.}
\label{tab:symbols}
\resizebox{\linewidth}{!}{%
\begin{tabular}{cl}
\hline
\textbf{Symbols} & \textbf{Definitions} \\
\hline
    $C_{i}$ & The total resources consumed by each microservice $i$ \\
    $\text{Tail}\_{\text{latency}_k}$ & The end-to-end tail latency of each chain $k$ \\
    $SLO_k$ & Set SLO constraints for each microservice chain $k$ \\
    $T_i$ & The service time for microservice $i$ in a chain \\
    $\tau$ & Zero-load latency of the microservice \\
 \color{black}   $\lambda$ & \color{black}{Arrival rate, typically defined as the number of requests arriving at the system per unit of time (per second)}  \color{black}\\
    $\mu$ & The maximum arrival rate that an instance of the microservice can sustain (without incurring infinite queuing)\\ 
    $\phi$ & The shape of the LLP graph \\
    $s_i$ & SLO resources allocated to each microservice $i$ \\
    $\sigma$ & The cost of a single instance of a microservice \\
    $ReC$ & The relative cost of a microservice \\
    $TC$ & The total cost for all microservice \\
    $pTL_i$ & Partial tail latency for each microservice $i$ \\
    $eTL$ & End-to-end tail latency \\
    $PSLO_i$ & Partial SLO resource for each microservice $i$ \\
    $x_i$ & Proportion of traffic flowing to each chain over a period of time $i$ \\
    $O$ & Oscillation factor \\
    $OCT$ & \textcolor{black}{Defined oscillation coefficient thresholds} \\
    $CL_i$ &  The load at a specific point in a time period \\
    $\bar{CL}$ &  The average load value over the time period \\
    $E_i$ & The set of SLO resource assigned to microservice $i$ during iterative computation \\
    $\bigtriangleup ReC_i$ & The change in relative cost per additional SLO block\\
    \textcolor{black}{$d$} & \textcolor{black}{ Appropriate number of SLO blocks (in this paper we defined as 1000)}\\
    \textcolor{black}{$F$} & \textcolor{black}{ The functional model of the $OCT$}\\
\hline
\end{tabular}%
}
\end{table}

GrandSLAm \cite{kannan2019grandslam} proposes to divide the end-to-end latency SLOs between microservices in proportion to their average service time (as shown in Eq.~\eqref{grand}), but this approach does not fully optimize the reservation consumption and failure to meet the QoS needs of users. 
\begin{equation}
SLO_k = \frac {T_k} {\sum \limits_{i=1}^{n} T_i},\label{grand}
\end{equation}
where $T_i$ denotes the service time for microservice $i$ in a chain.

\begin{figure}[htbp]
\centering
\begin{minipage}[b]{0.49\linewidth}
  \centering
  \includegraphics[width=\linewidth]{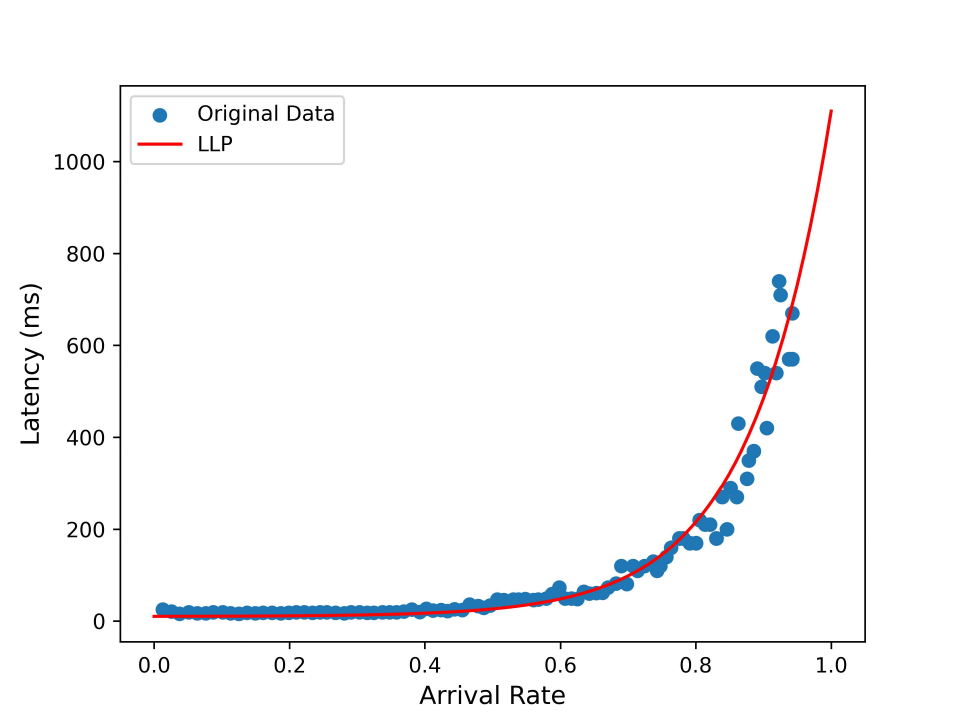}
  \caption{LLP Graph.}
  \label{fig:LLP}
\end{minipage}
\begin{minipage}[b]{0.49\linewidth}
  \centering
  \includegraphics[width=\linewidth]{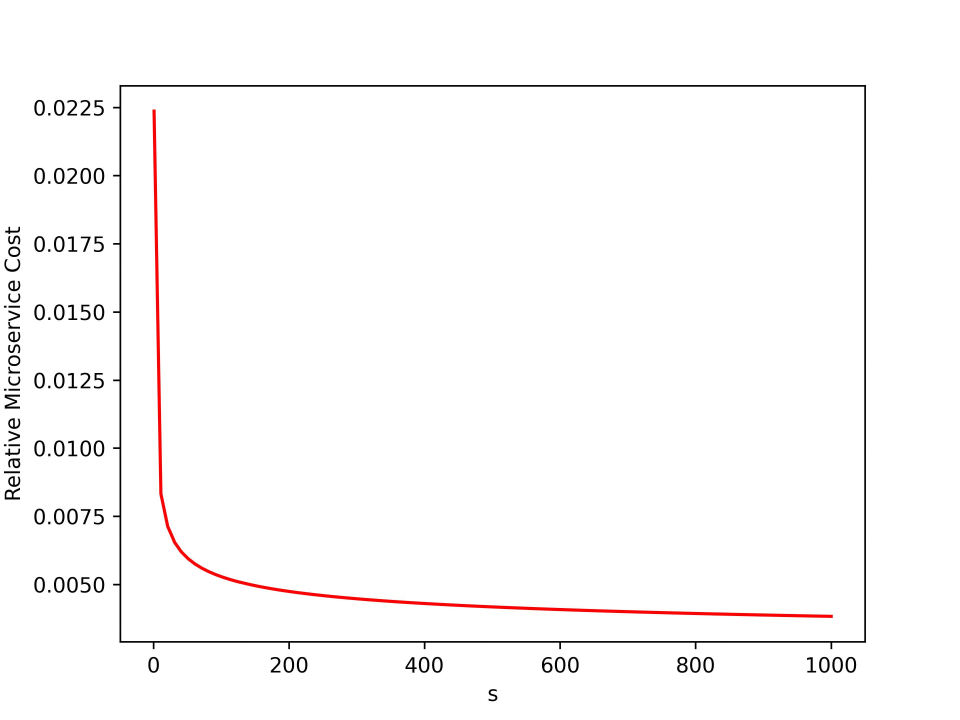}
  \caption{\textcolor{black}{Relative Microservice Cost.}}
  \label{fig:Cost}
\end{minipage}
\end{figure}

Parslo proposes, based on the LLP graph (shown in Figure.~\ref{fig:LLP}, with the function expression shown in Eq.~\eqref{response}) of microservice and the above model, the relationship between the relative total cost of microservice resource consumption and the allocation of the SLO resources to each microservice (as shown in Eq.~\eqref{rela_cost} and Figure.~\ref{fig:Cost}). \textcolor{black}{We define the 
$\sigma$ parameter as the cost of a single microservice instance (like a pod in Kubernetes). We assume that all instances of a single microservice have the same cost without auto-scaling to change resources. When all microservices are deployed on the same type of instance, $\sigma$ can be set to 1. However, since each microservice may be deployed on instances with different configurations (e.g., CPU cores, memory) or even different hardware capabilities, it is more realistic to optimize for the total cost of the microservices rather than simply considering the total number of instances. In the Figure.~\ref{fig:Cost}, $s$ represents the SLO resources allocated to the microservice.}
\begin{equation}
R = \tau \phi \left( \frac{\lambda}{\mu} \right), \label{response}
\end{equation}
where $R$ denotes the RT of the microservice request, $\tau$ is the  zero-load latency of the microservice, $\mu$ specifies the highest arrival rate that a microservice instance can handle without resulting in infinite queuing, $\lambda$ is an arbitrary arrival rate, and the $\phi$ function is the the shape of LLP graph of a microservice instance. 
\begin{equation}
ReC = \frac {\sigma}  {\mu \phi^{-1}(\frac {s}{\tau})}, \label{rela_cost}
\end{equation}
where $ReC$ specifies the relative cost of a microservice, $\sigma$ denotes the cost of a single instance of a microservice.

For a single chain with multiple microservices, Parslo assumes that when the SLO latency uses the end-to-end average latency as a metric, Eq.~\eqref{total_cost} is used as a scale for SLO resource allocation to find the optimal SLO resource allocation scheme for each microservice, resulting in the least overall resource consumption.
\begin{equation}
TC  = \sum_{i=1}^n ReC_i = \sum_{i=1}^n \frac{\sigma_i}{\mu_i \phi_{i}^{-1}(\frac {s_i}{\tau_i})},\ with \ \sum_{i=1}^n s_i = SLO, \label{total_cost}
\end{equation}
where $TC$ denotes the total cost for all microservice.

\begin{figure*}[htbp]
\centering
\includegraphics[width=1\linewidth]{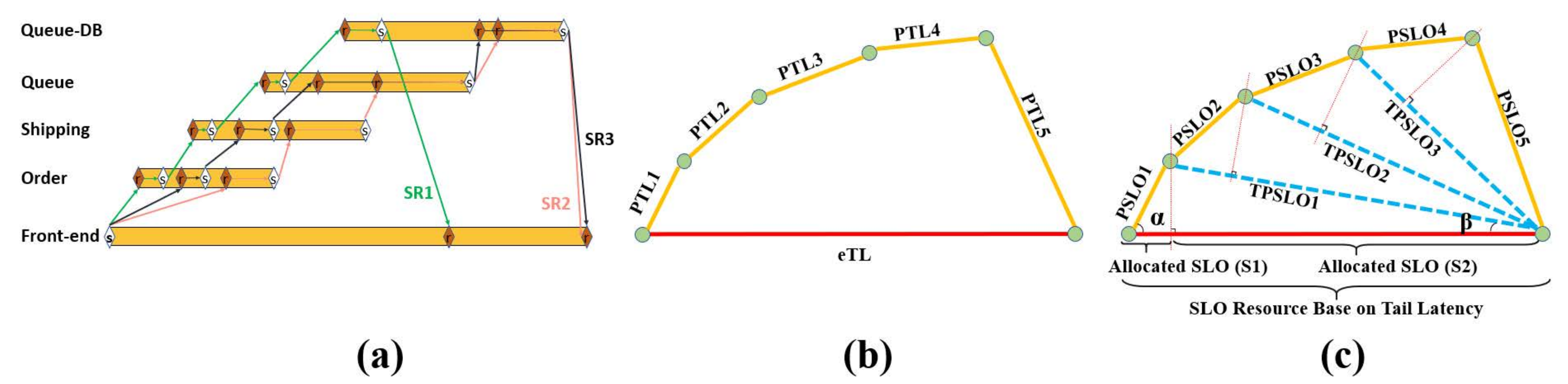}
\caption{Lightweight SLO Allocation Model.}
\label{fig:light_model}
\end{figure*}

\color{black} However, when the end-to-end average latency is adopted as the SLO metric, it fails to fully capture the user's overall service satisfaction. Therefore, selecting the eTL as the SLO metric becomes necessary to ensure comprehensive user satisfaction. Nevertheless, using tail latency as the end-to-end SLO for resource allocation results in resource wastage, primarily due to tail latency being a non-accumulative metric.\color{black}
We have characterized the local tail latency incurred by different components processing service requests (SRs) on a microservice chain, as illustrated in Figure.~\ref{fig:light_model}a. Here, SR1 represents the fastest service request processed among all requests,  \textcolor{black}{which also represents an optimal situation, showing the shortest time for each microservice to handle the requests}. SR2 represents the slowest case of requests processing, with the longest processing time in each microservice component, serving to characterize the end of the tail latency for each component. SR3 represents the majority of requests contributing to end-to-end latency, where service requests are processed at normal speeds in some components but encounter longer processing times in one component, resulting in them being the slowest batch in terms of end-to-end latency.
The quantitative relationship between these aspects is provided by Eq.~\eqref{ptl}. The sum of partial tail latency (PTL) exceeds the eTL, leading to an excess of resources allocated to each microservice when eTL is utilized as the total SLO resource for allocation. Parslo proposes a combination of offline and online analytical models for addressing this issue, requiring significant time and computational resources for fitting in order to achieve a more satisfactory SLO resource allocation result. 
\begin{equation}
\sum_{i=1}^n {pTL_i} \geq eTL . \label{ptl}
\end{equation}

\textcolor{black}{To reduce the consumption of computational and time resources and quickly obtain the optimized SLO resource allocation scheme, we model the relationship between the PTL for each microservice on a single chain and the eTL, as shown in Figure.~\ref{fig:light_model}b. The relationship between the \textcolor{black}{$\vec{PTL}$} and \textcolor{black}{$\vec{eTL}$} as a sum of vectors as shown in Eq.~\eqref{etl}:}
\begin{equation}
\sum_{i=1}^n {\vec{PTL_i}} = \vec{eTL}. \label{etl}
\end{equation}

Based on this abstract model, we use the \textcolor{black}{$\vec{eTL}$} as a resource metric for SLO resource allocation, and compute the allocation by the gradient descent method combined with the LLP of microservices to get the allocated SLO.

\textcolor{black}{Through iterative optimization using gradient descent, we can find a nearly optimal partial SLO allocation scheme, thereby minimizing the deployment cost of end-to-end services. Although gradient descent typically guarantees convergence to a local optimum, as shown in Eq.~\eqref{total_cost}, the cost function we aim to optimize is convex. Therefore, we can use the gradient descent method to find the global optimal solution.}

\textcolor{black}{According to this, we construct vectors to derive Partial SLO (\textcolor{black}{$\vec{PSLO}$}), the value of which is defined in Eq.~\eqref{PSLO} and closely approximates the actual \textcolor{black}{$\vec{eTL}$}. To simplify the calculation process, we consolidate the latter nodes into one node (referred to as typical partial SLO, \textcolor{black}{$\vec{TPSLO}$}), as depicted in Figure.~\ref{fig:light_model}c. However, the allocation process involves computing the SLO resources allocated to each node separately, ensuring that it does not impact the results of the SLO resource allocation.}
\begin{equation}
\vec{PSLO_1} = \frac {s_1} {\cos \alpha} \ , \ \vec{TPSLO_1} = \frac {s_2} {\cos \beta},\label{PSLO}
\end{equation}
where $\vec{PSLO_1}$ indicates the amount of SLO resources to be allocated to microservice 1, $s_1$ denotes the assigned SLO value based on the \textcolor{black}{eTL} SLO calculation, $\vec{TPSLO_1}$ means the sum of SLO resource vectors after $\vec{PSLO_1}$.

However, there exists no fixed relationship between pTL and eTL. Different types of microservices may exhibit various correlations, influenced by factors such as cluster state, internal communication, and hardware conditions. The mapping to the value of $PSLO$ cannot be definitively determined, manifesting itself in the form of the angle between the vectors $\vec{PSLO_1}$ and $\vec{PSLO_1}$. 

\color{black} In order to maintain our lightweight and fast computation objective, we further simplify the model by setting the angle between the vectors $PSLO_1$ and $TPSLO_1$ to be a right angle. The point of intersection between them lies on the perpendicular line where the SLO assignment point is located. Consequently, we derive a lightweight solution to determine the PSLO values based on eTL SLO assignments, as shown in Eq.~\eqref{PSLO_fi}  to Eq.~\eqref{PSLO_s}: \color{black}   
\begin{equation}
PSLO_i = \sqrt{SLO_{end-to-end}*s_i}, \label{PSLO_fi}
\end{equation}
\begin{equation}
s_i = s_0 + \sum_{b_j\in E_i} b_j ,
\end{equation}
\begin{equation}
E_i = \{b_1...b_j...b_m\}
\left\{
\begin{aligned}
E_1 \cup E_2 \cup...\cup E_i \cup ...\cup E_n = E  \\
E_1 \cap E_2 \cap...\cap E_i \cap ...\cap E_n = \varnothing \\
\end{aligned}
\right. ,\label{PSLO_s}
\end{equation}
where $s_i$ denotes the SLO resource allocated to each microservice $i$, $s_0$ represents a fixed value that is pre-assigned to each microservice, $b_j$ denotes which round of iteration allocates SLO resources to microservice $i$, $E_i$ indicates the set of SLO resource assigned to microservice $i$ during iterative computation, $E$ denotes the total set of SLO resource iteration allocations.

In order to achieve the goal of lightweight model and facilitate real-time computation, compared with the scheme of fitting to meet the optimal SLO resource allocation by constantly modifying the SLO resources, we choose this lightweight computation to obtain the optimized scheme, which can ensure the rapid adaptation of LSRAM in the environment of constant transformation of microservice applications and high-speed change of loads, which can not only ensure the QoS of the users, but also save the resources.

\begin{figure}[htbp]
\centering
\includegraphics[width=\linewidth]{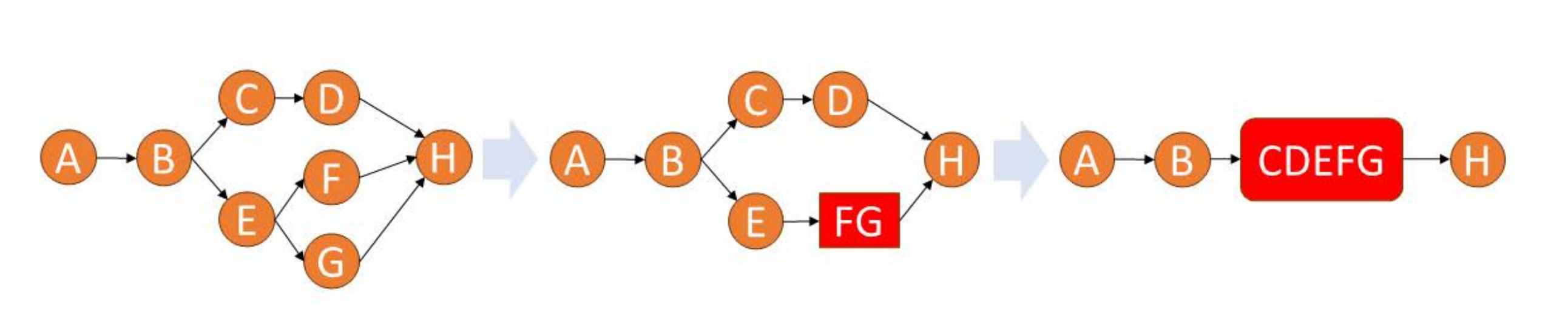}
\caption{Chains Merge.}
\label{fig:chain}
\end{figure}

\subsection{SLO Resource of Multi-chain Shared Microservices and SLO Resources Update Algorithm} \label{allocate}

\textcolor{black}{For complex chains that have typical structures such as dynamic branching, parallel fan-out, and microservice dependencies, similar to the end-to-end chains of Nested Fork Join DAGs\textcolor{black}{~\cite{NFJDAG}}, we adopt the approach of continuous merge into a single chain for calculation, so that each microservice chain with the same chain endpoint is finally merge into a single chain, in which the complex chain structure is abstracted into a microservice node (the merge process is shown in Figure.~\ref{fig:chain}). }

\begin{algorithm}
    \caption{SLO Resource Initialization Allocation Algorithm.}
    \label{alg:initial}
    \KwIn{SLO constraints for each microservice chain $SLO_{k}$, microservice load-latency-profile $\tau \phi \left( \frac{\lambda}{\mu} \right)$, microservice chain relationships $L$.}
    \KwOut{SLO resources allocated to each microservice $s_i$}

    \textbf{Initialize parameters:}\\
    \ \ SLO resource $SLO_{resource}$, appropriate number of blocks $d$, number of microservice $n$ in a chain, predefined SLO resource segments $SLO_{block}$, number of chains $k$, SLO resource $SLO_{shared}$ assigned to multi-chain shared microservices, the amount of change in ReC $\bigtriangleup ReC_i$ after each additional SLO block. \\
    \textbf{SLO Allocation:} \\
    $SLO_{block} = \frac{SLO_k}{d}$ \\
    $SLO_{resource} = SLO_{k} - (n * SLO_{block} )$ \\
    \If{$SLO_{resource} \geq SLO_{block}$}{
    $\bigtriangleup ReC_i = \frac{\sigma_i}{\mu_i \phi_{i}^{-1}(\frac {s_i}{\tau_i})} - \frac{\sigma_i}{\mu_i \phi_{i}^{-1}(\frac {s_i + SLO_{block}}{\tau_i})}$ \\
    Find the microservice with the largest $\bigtriangleup ReC_i$ in the queue\;
    Add SLO resources to this microservice\;
    $s_{i} = s_{i} + SLO_{block}$ \\
    $SLO_{resource} = SLO_{resource} - SLO_{block}$\\
    }
    $SLO_{shared} = \frac{\sum\limits_{j=1}^k {SLO_j}}{k}$
\end{algorithm}

During the resource allocation process, we partition the SLO resources into several equally-sized blocks. The size of these blocks must strike a balance, they should not be too large, as this could hinder the search for an optimal allocation scheme, nor too small, as this might excessively prolong the computation time. \textcolor{black}{To balance the trade-offs, we allocate SLO resources ranging from 800 to 1200 SLO blocks, and each block is around 4ms.} Before the initial allocation phase, we distribute a specific number of SLO resource blocks to each microservice to ensure that, when employing the gradient descent method, it encounters a convergence point. This strategy helps prevent the allocation algorithm from continuously allocating resources to the same microservice indefinitely. Typically, LSRAM pre-allocates 10 SLO resource blocks \color{black}(e.g. 40ms) \color{black}to each microservice. We then calculate the relative increase in resource consumption for each microservice along the single chain after adding a resource block. The change in resource consumption following the addition of each SLO block is illustrated in Equation \eqref{fenpei}. \textcolor{black}{If there are cases where the SLO resources cannot be evenly divided by the SLO blocks, due to the large number of blocks we use, only a very small amount of SLO resources will generally remain. These remaining SLO resources will be allocated together to the microservice that saves the most resources during the final iteration.}

\begin{equation}
\bigtriangleup ReC_i = \frac{\sigma_i}{\mu_i \phi_{i}^{-1}(\frac {s_i}{\tau_i})} - \frac{\sigma_i}{\mu_i\phi_{i}^{-1}(\frac {s_i + SLO_{block}}{\tau_i})}, \label{fenpei}
\end{equation}
where $\bigtriangleup ReC_i$ denotes the change in relative cost per additional SLO block.

After each microservice on a single chain increases by one resource block, we compute its relative increase in resource consumption. Subsequently, we select the microservice with the least resource consumption. The evaluation criterion for the abstract node is the point with the least resource consumption increment in its status. The specific SLO resource allocation algorithm for the terminal of the same chain is depicted in Algorithm~\ref{alg:initial}, where the SLO resource value assigned to the abstract node will actually be assigned to the point with the least incremental resource consumption within it.

In Algorithm~\ref{alg:initial}, the inputs of the algorithm include $SLO_{k}$ (end-to-end SLO constraints), the LLP graph function $\tau \phi \left( \frac{\lambda}{\mu} \right)$ for each microservice in a chain, and the relationship between the microservices in the chain (lines 1-3). Line 6 are the parameter definitions. To expedite computation, the algorithm divides the SLO resources into a number of $SLO_{block}$ (line 8), which are then allocated one by one to the most resource-efficient microservices. Before starting the iterative allocation of SLO resources, it is necessary to allocate some $SLO_{block}$ to each microservice in advance, crossing the point of the extreme value of $ReC_i$, to ensure that there is no negative number in the computation of $\bigtriangleup ReC_i$ (line 11). We utilize a queue to store the resource consumption of each microservice each time an $SLO_{block}$ is allocated (line 12), ensuring that each update only involves the microservice node with the allocated SLO resources. Following SLO resource allocation, we initialize the SLO resources of the shared microservice (line 12).

\begin{figure}[htbp]
\centering
\includegraphics[width=\linewidth]{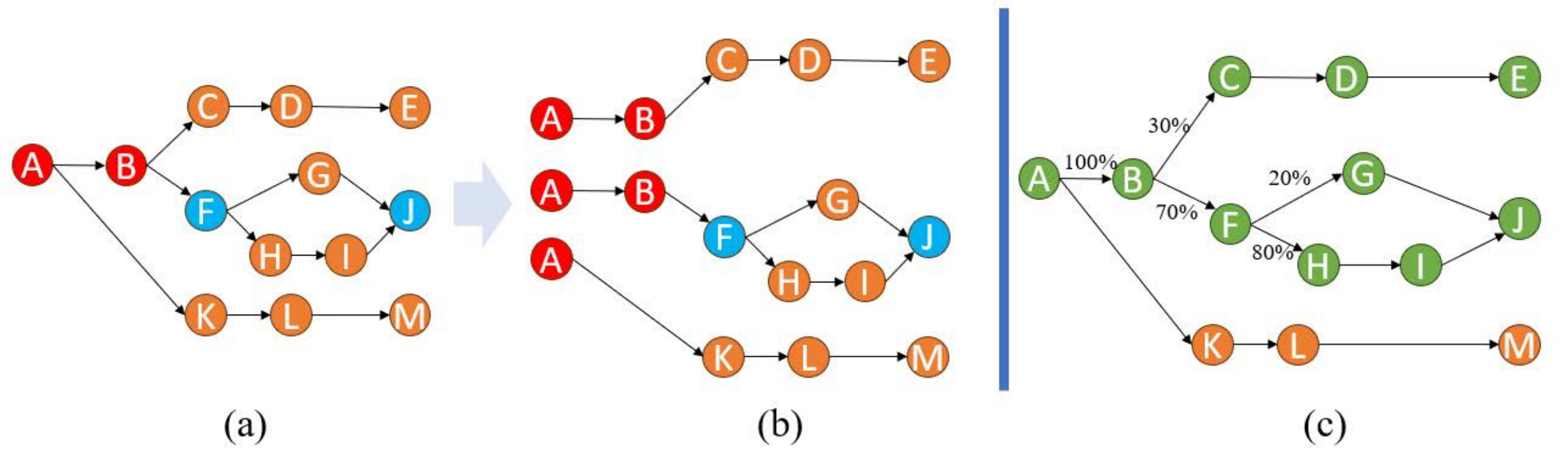}
\caption{(a) denotes the chains structure of the microservice; (b) indicates the splitting of the microservice chain architecture to obtain multiple single chains; (c) specifies proportion of traffic going to different chains over a period of time.}
\label{fig:chain2}
\end{figure}

\textbf{Complexity Analysis:} The time complexity of Algorithm~\ref{alg:initial} is $O(n\log n)$, where $n$ represents the number of $SLO_{block}$, \textcolor{black}{as in the algorithm, we use a queue to store the $\bigtriangleup ReC_i$ values after each calculation. In each iteration, only the $SLO_{block}$ value of the microservice with the highest value from the previous round needs to be updated. The entire calculation process is only related to the number of $SLO_{blocks}$. Once the $SLO_{blocks}$ have been distributed, the initialization of the entire SLO resources is completed, this meets our model's requirements for lightweight and fast computation.} 

However, managing microservices that are shared across multiple chains presents a challenge in determining the appropriate SLO resource allocation for each chain. This complexity arises in selecting the SLO resource value to use as a resource scheduling metric for the shared microservice. To overcome this problem, we conduct a detailed analysis of scenarios involving such shared microservices. In the initial allocation phase, we treat the shared microservice as unique and independent for each chain, as depicted in Figure.~\ref{fig:chain2} to illustrate replication separation. Based on this configuration, we compute the SLO resources allocated to the shared microservice at each chain endpoint. During the initial allocation process, we adopt the method showed in Equation \eqref{10} to obtain an initial allocation value for the SLO resources.

\begin{equation}
SLO_{shared} = \frac { \sum \limits_{j=1}^k {SLO_j} }  {k}, \label{10}
\end{equation}
\begin{equation}
SLO_{shared} = \sum \limits_{j=1}^k { x_{j} * SLO_{j} }, \label{11}
\end{equation}
where $k$ denotes the number of microservices in a chain, $x_{j}$ denotes the proportion of traffic for any microservice in a chain, $SLO_{shared}$ indicates the SLO resources that should be allocated to the multi-chain shared microservice.

\begin{algorithm}[h]
    \caption{SLO Resource Update Allocation Algorithm.}
    \label{alg:update}
    \KwIn{ Proportion of traffic distributed to each chain over a period of time $x_i$, initial allocation of SLO resources to each microservice $s_i$, microservice chain relationships $L$.}
    \KwOut{SLO resources allocated to each microservice $s_i$}

    \textbf{Initialize parameters:}
    \ \ SLO resource $SLO_{shared}$ assigned to multi-chain shared microservice, elapsed time $t$, predefined resource reallocation time thresholds $M_t$, SLO resource $SLO_{new}$ of new microservice.\\
    \textbf{Upedate:}\\
    \If{$t \geq M_t$ or $cluster \  environment \  changes$}{
        $SLO_{shared} = \sum \limits_{j=1}^k {x_{j}*SLO_{j}}$ \\
        Reallocating SLO resources for microservices $s_i$ \\
        \If{$cluster \  environment \  changes$}{
            Collect the LLP graphs for new microservices $i$ \\
            $s_i = s_0 + \sum_{b_j\in E_i} b_j$ \\
            \If{New microservie connects to shared microservice}{
            $SLO_{shared} = SLO_{shared} + \frac{1}{k} SLO_{new}$}
        }
    }
\end{algorithm}

In the actual production environment, the loads distributed to different chains within a period of time are inconsistent Figure.~\ref{fig:chain2}c, in order to ensure that our model can continuously adapt to different production environments, we design an updated algorithm for LSRAM to allocate SLO resources for shared microservices, which lies in the fact that the LSRAM collects the load traffic flow of the shared microservices flowing to different end-to-end services within a period of time. LSRAM's SLO resource allocation update algorithm still follows the principle of lightweight, and can adapt to changes in the load flow direction of the cluster Eq.~\eqref{11}, so that the cluster scheduling can ensure user service satisfaction. LSRAM's SLO resource update allocation algorithm still follows the principle of lightweight and adapts to changes in the load flow of the cluster, enabling cluster scheduling to ensure user service satisfaction. When the microservice application, attribute, or instance specification in the cluster changes, LSRAM will reallocate the microservice's SLO resources. Our SLO resource update allocation algorithm is shown in Algorithm~\ref{alg:update}. Lines 1 to 3 are the inputs requested by the algorithm, line 5 is the settings of the parameters, when there is a change in the cluster environment (\textcolor{black}{including changes in system resource capacity, changes in cluster hardware, redeployment of containers, etc.}) or a pre-set time is reached, the algorithm will reallocate the SLO resources of the microservices (line 10) and compute the SLO resources of the shared microservices according to line 11.

\subsection{esDNN Prediction Model and Mode Converter}

To enhance the LSRAM's ability to cope with highly fluctuating loads and bursty loads, we utilize the esDNN~\cite{xu2022esdnn} load prediction model and design the mode converter. \textcolor{black}{The esDNN model is our previous work which is an improved model based on Gated Recurrent Unit (GRU). It's lightweight and straightforward, enabling quick predictions of the load value for the next time slot}. 

We trained our prediction model using Alibaba's dataset \cite{alibaba2022} for updating, focusing on high volatile changes in load and burst load for intensive training. The training was also targeted on the load of different microservices, which ensures the accuracy of the prediction model in predicting the load of different microservices for the next time slice.



\begin{figure}[htbp]
\centering
\begin{subfigure}[t]{0.45\linewidth}
  \centering
  \includegraphics[width=\linewidth]{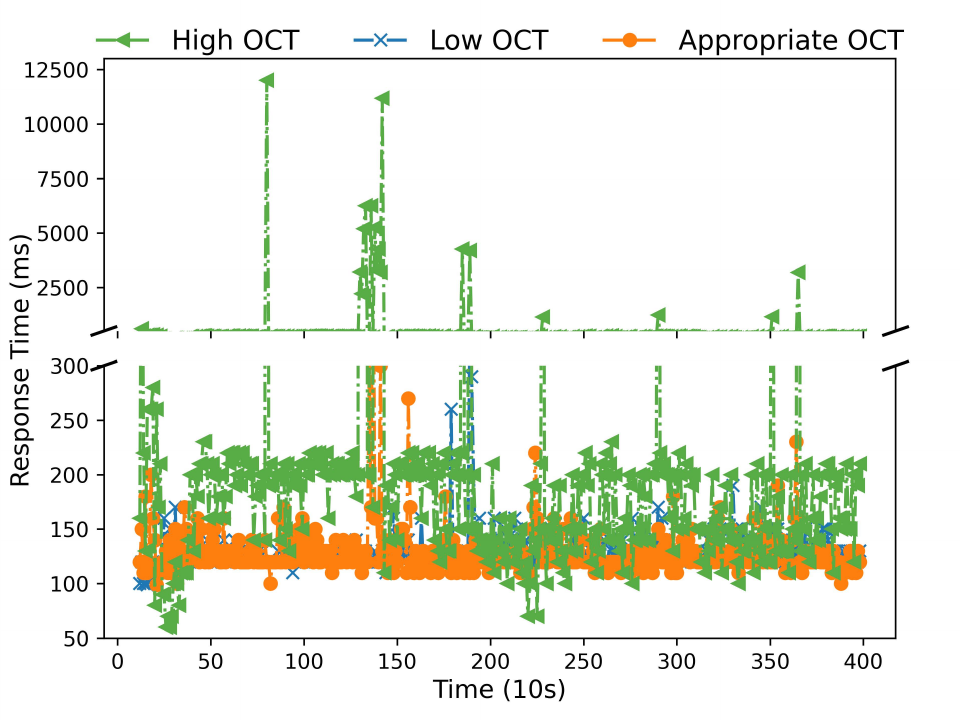}
  \caption{\textcolor{black}{The impact of different OCTs on response time under the same load conditions.}}
  \label{fig:diffoc}
\end{subfigure}
\begin{subfigure}[t]{0.48\linewidth}
  \centering
  \includegraphics[width=\linewidth]{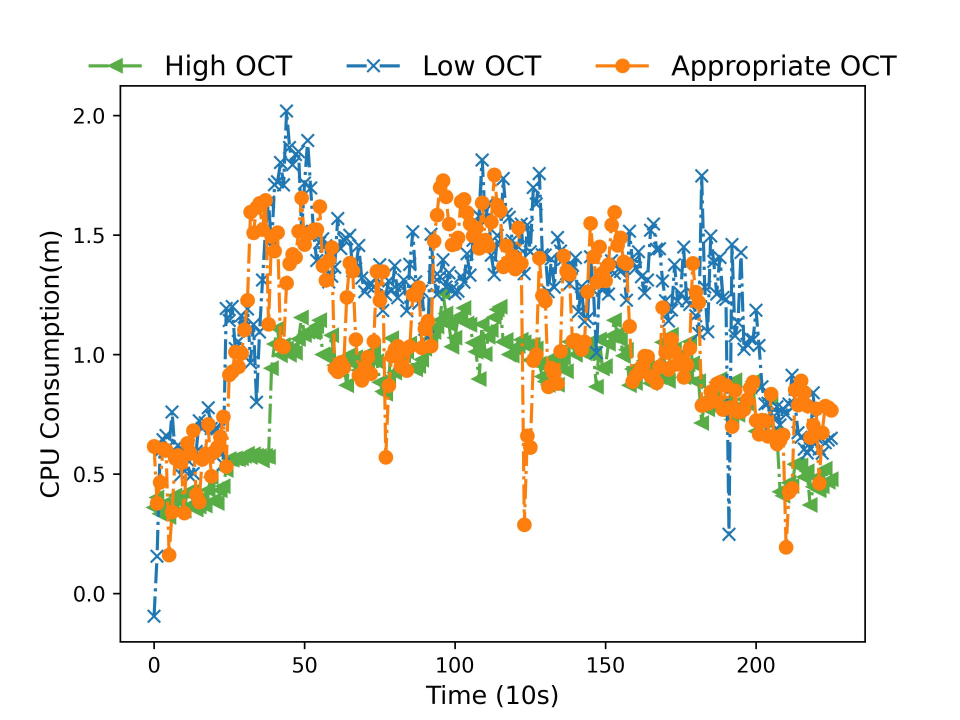}
  \caption{\textcolor{black}{CPU consumption of microservices under different OCTs.}}
  \label{fig:cpuoc}
\end{subfigure}
\caption{\textcolor{black}{Performance under different OCTs.}}
\label{fig:oc}
\end{figure}


To ensure the QoS of users to the maximum extent under high fluctuating loads, we design a mode converter for high fluctuating loads, and when the oscillation-aware function in the mode converter detects that the system is in a state of high fluctuating loads, it will change the resource scaling policy of the resource scheduler to ensure the QoS of users. \textcolor{black}{We use a sliding window mechanism to store past load information and detect if the load is oscillating by calculating the variance of the load relative to the average load. A high $O$ value indicates significant load fluctuations, while a low $O$ value indicates more stable load conditions.}
The oscillation-aware function is shown in Eq.~\eqref{oscillation-aware}:
\begin{equation}
O = \frac{\sqrt{\frac{1}{n}\sum \limits_{i=1}^{n}[CL_{i}-\bar{CL}]^2}}{\bar{CL}} , \label{oscillation-aware}
\end{equation}
where $O$ denotes the oscillation factor, $CL_{i}$ denotes the load at a specific point in a time period, $\bar{CL}$ indicate the average load value over the time period.

\textcolor{black}{An appropriate oscillation coefficient threshold (OCT) is crucial for the mode converter. If the oscillation coefficient is set too high, the converter cannot detect system fluctuations, failing to switch modes promptly, which results in severe SLO violations in microservices. Conversely, if the oscillation coefficient is set too low, the system frequently switches to a high oscillation mode, leading to resource wastage. As illustrated in the Figure.~\ref{fig:oc}, we have demonstrated the impact of different oscillation coefficient thresholds on response times and resource utilization. We can observe that when the $OCT$ is set too high, the system becomes insensitive to fluctuating loads, resulting in serious SLO violations. On the other hand, when the $OCT$ is set too low, the system becomes overly sensitive to load fluctuations, causing the frequent resource scaling operations. Although this effectively ensures service quality, it leads to significant resource waste.}

\textcolor{black}{Moreover, different microservices exhibit varying sensitivities to load fluctuations, so the $OCT$ should be set accordingly for each microservice. Through analysis and synthesis, we found that the $OCT$ setting is related to the LLP parameters of the microservices and the partial SLO assigned to them. The steeper the LLP curve of a microservice, the more sensitive it is to load changes. When the same number of requests is sent to microservices with different LLP steepness, the steeper ones experience a faster increase in response time, making them prone to request timeouts and service crashes when receiving fluctuating loads. Additionally, the more partial SLO assigned to a microservice, the more susceptible it becomes to fluctuating loads, meaning that it requires more SLO time to process the same number of requests. Based on this, we modeled the relationship between these three factors and derived the Eq.~\eqref{oscillation-llp}.}
\begin{equation}
\color{black}
OCT = (F(\frac{{(PSLO)}^2}{\mu\phi^{-1}\left(\frac{PSLO}{\tau}\right)})),
\color{black}
\label{oscillation-llp}
\end{equation}
\color{black}where $F$ denotes the functional model of the $OCT$. \color{black} 

\begin{figure}[htbp]
\centering
\begin{minipage}[t]{0.45\linewidth}
  \centering
  \includegraphics[width=\linewidth]{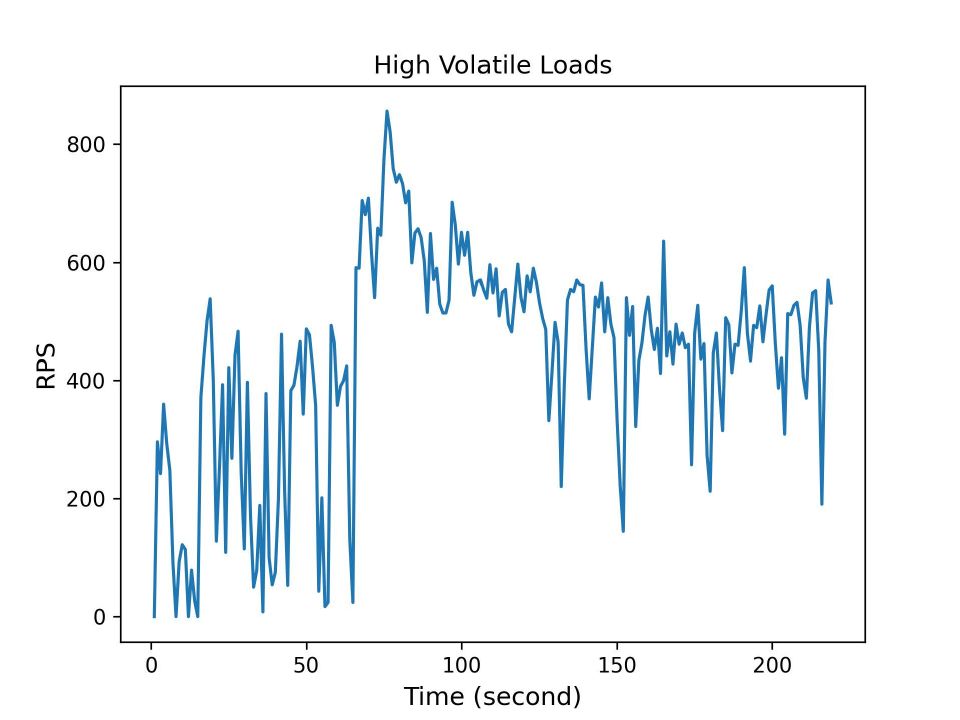}
  \caption{\textcolor{black}{The highly volatile load.}}
  \label{fig:llp4}
\end{minipage}
\begin{minipage}[t]{0.45\linewidth}
  \centering
  \includegraphics[width=\linewidth]{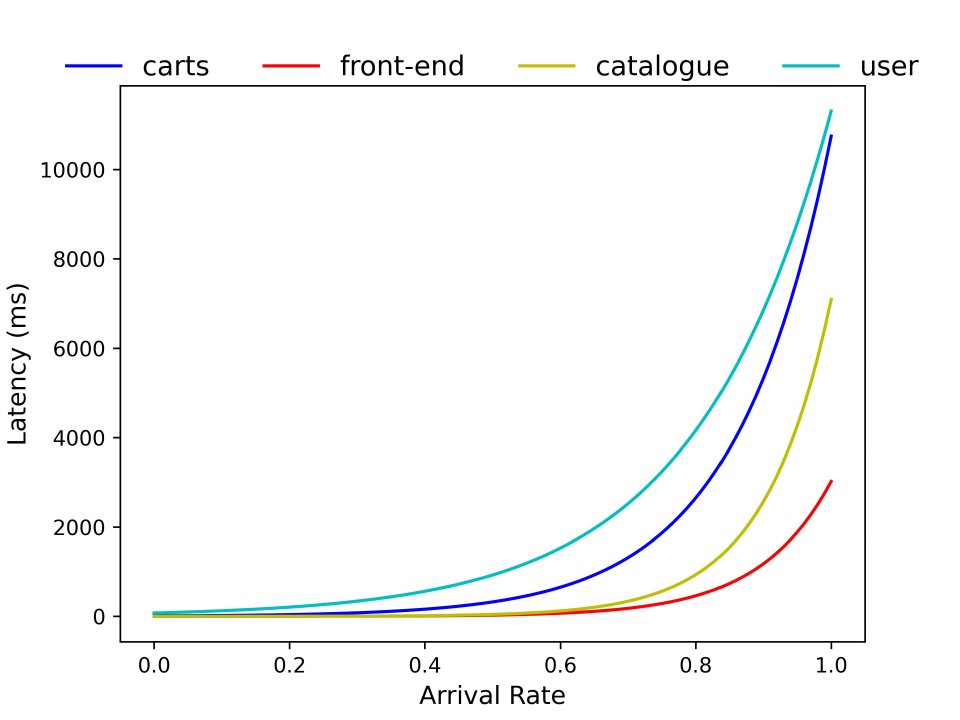}
  \caption{\textcolor{black}{LLP graphs of Sock-Shop's critical microservices.}}
  \label{fig:high}
\end{minipage}
\end{figure}

\section{EXPERIMENTAL EVALUATIONS} \label{experimental}


In this section, we present information about our experiments, including the environment setup for the experiment, the datasets used and the application. Secondly, we will demonstrate and analyze the results of our experiments and explain the improvement of our methodology.

\subsection{Experimental Setup}

\textbf{Cluster Configuration}: All of our experiments are validated on a prototype system with following server configurations: there are a total of 10 nodes, comprising 3 master nodes and 7 worker nodes. 
\textcolor{black}{The 3 master nodes are equipped with 32 cores and 64GB memory each, among the 7 worker nodes, 4 are configured with 32 cores and 64GB memory, and the remaining worker nodes are with 56 cores and 128GB memory, 104 cores and 256GB memory, 64 cores and 64GB memory. The microservices were deployed on top of Kubernetes.}

\textbf{Datasets}: To verify the effectiveness of our method in a realistic environment, we employed the dataset obtained from Alibaba Cluster \cite{alibaba2022}, which was sourced from production clusters within Alibaba encompassing over ten thousand bare-metal nodes during a 13-day period in 2022. The dataset records the CPU and memory utilization of over 470,000 containers spanning more than 28,000 microservices within the same production cluster. We transform the data from this dataset into request counts and generate workloads using Locust\footnote{https://locust.io/}, which was configured to mimic application requests for stress tests. \textcolor{black}{We also convert the fluctuations in CPU utilization on the server to changes in the number of requests sent by Locust by measuring the relationship between container CPU utilization and load requests in the Alibaba dataset}.


\begin{figure}[htbp]
\centering
\begin{subfigure}{\textwidth}
  \centering
  \includegraphics[width=0.76\linewidth]{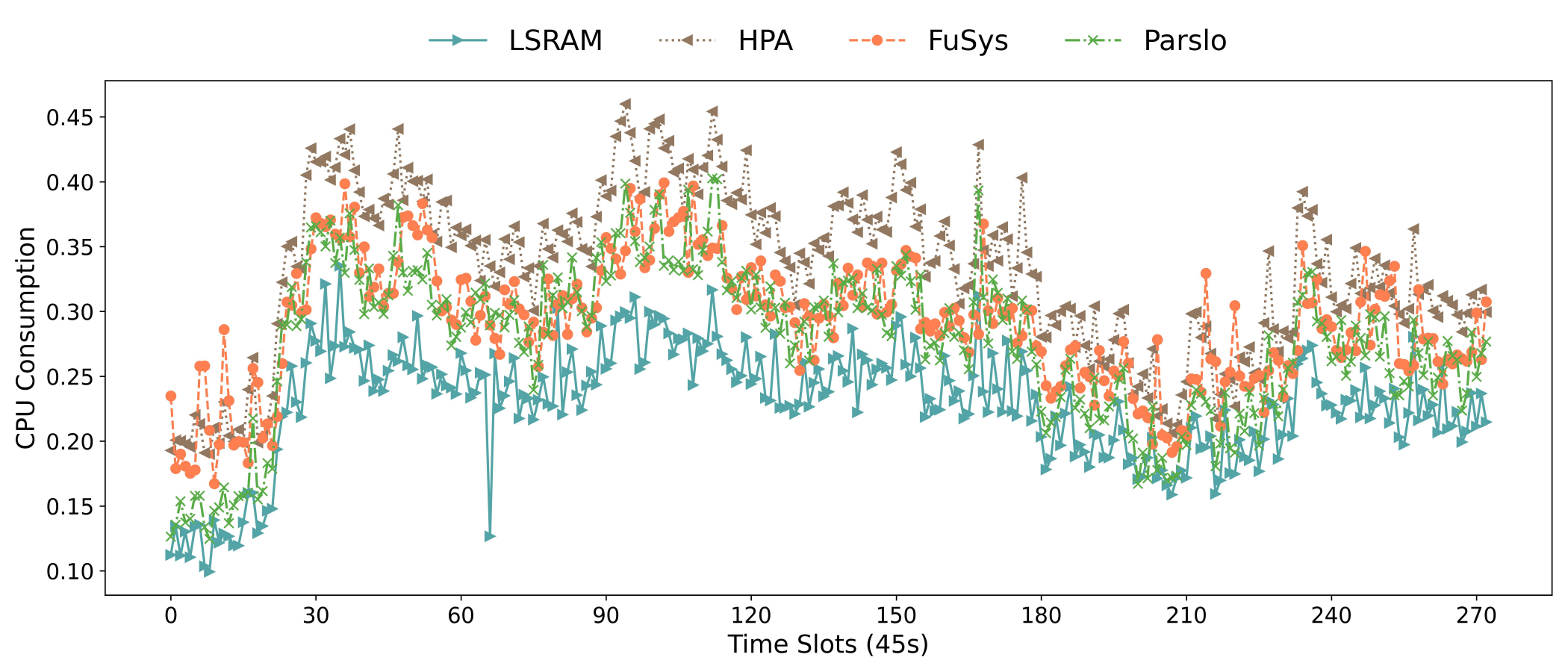}
  \caption{CPU Resource Consumption.}
  \label{fig:cpu}
\end{subfigure}
\begin{subfigure}{\textwidth}
  \centering
  \includegraphics[width=0.76\linewidth]{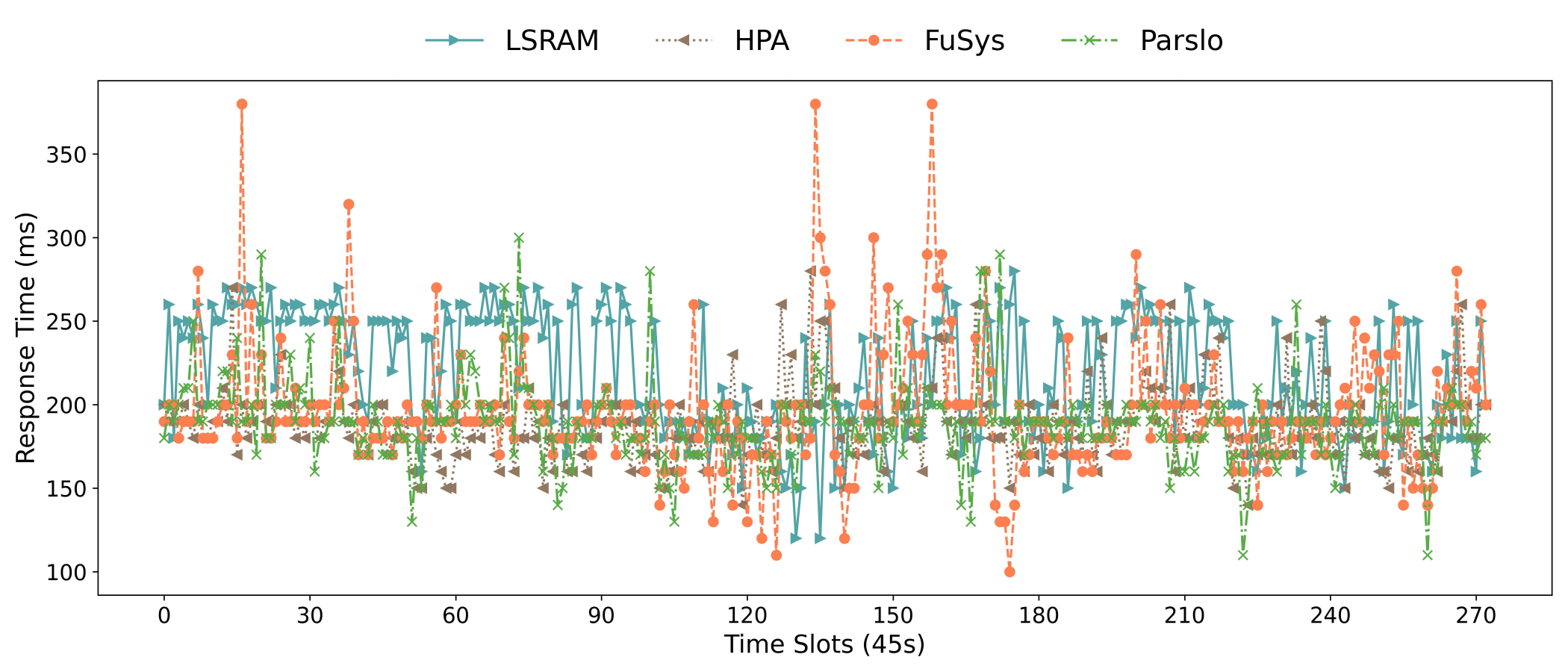}
  \caption{Response Time (P99).}
  \label{fig:response}
\end{subfigure}
\caption{Performance under the Same Load.}
\label{fig:result1}
\end{figure}

\textbf{Tested Application}: Sock-shop \cite{sockshopdemo} is an e-commerce website microservice whose functions include searching, ordering, and shipping, which can be divided into three parts: front-end, business logic, and database, and contains a total of 13 microservices. We conducted breadth load testing on the microservice components of Sock-shop's critical chains and carved out its LLP graph in detail, as shown in Figure.~\ref{fig:high}.

\subsection{Baseline Approach}

To evaluate the performance of LSRAM, we compare it with several state-of-the-art baselines as follows:

\textbf{Horizontal Pod Autoscaling (HPA)} \cite{qu2018auto}: It is a heuristic autoscaling framework based on a predefined threshold, which uses CPU utilization as its resource scaling metric in this experiment, and automatically scales resources when the system detects that the CPU utilization of an instance exceeds a fixed threshold. The threshold is set to [30\%, 50\%] in this experiment, as this value has been used in \cite{ahmad2024smart}.

\textbf{Fuzzy-Based Auto-scaler (FuSys)} \cite{liu2018fuzzy}: It is a heuristic dynamic threshold scheduling framework based on fuzzy control theory, Fuzzy will dynamically adjust the system's resource scheduling thresholds based on the system's current resource usage and load by fuzzy control.

\textbf{Parslo} \cite{mirhosseini2021parslo}: It is a method to compute the near optimal SLO resource allocation scheme based on the gradient descent method, each microservice will perform heuristic resource scaling based on this scheme.


\subsection{Evaluations}

We evaluate the proposed LSRAM with the above baselines in the same experimental settings. To avoid the randomness of experiments, we repeat each baseline 5 times and compute its average value to compare performance.

\textcolor{black}{The tests comparing LSRAM with other baselines clearly demonstrate its significant improvement on resource savings, as depicted in Figure.~\ref{fig:result1}a. LSRAM consistently consumes the minimum amount of system resources under the same load pressure variations. It maintains a relatively stable RT while minimizing CPU resource consumption. The scaling threshold of HPA was set with [30\%, 50\%] CPU utilization to ensure RT, resulting in the highest resource consumption. FuSys saves resources compared with HPA but experienced significant RT fluctuations. The reason why LSRAM and Parslo have relatively similar resource consumption in the early stages is that they both have near-optimal initial SLO resource allocation schemes. However, as the system keeps running, LSRAM continuously adapted to the current environment through its SLO resource updating algorithm, leading to a reduction in overall resource consumption.}

\begin{figure}[htbp]
\centering
\begin{subfigure}[b]{0.4\linewidth}
  \centering
  \includegraphics[width=\linewidth]{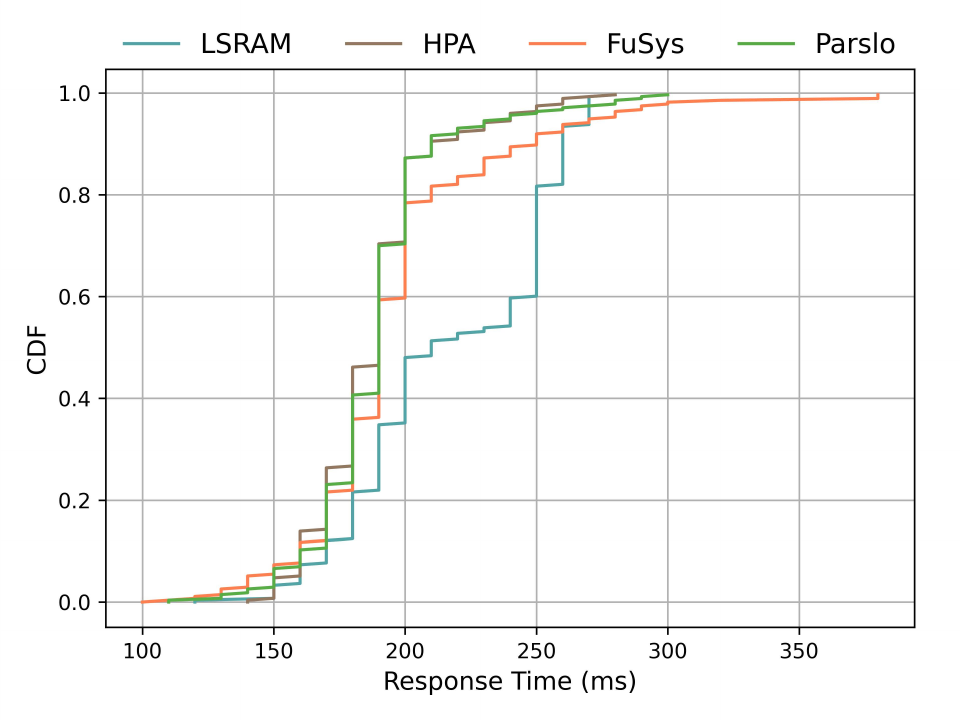}
  \caption{CDF for Response Time.}
  \label{fig:cdf}
\end{subfigure}
\begin{subfigure}[b]{0.4\linewidth}
  \centering
  \includegraphics[width=\linewidth]{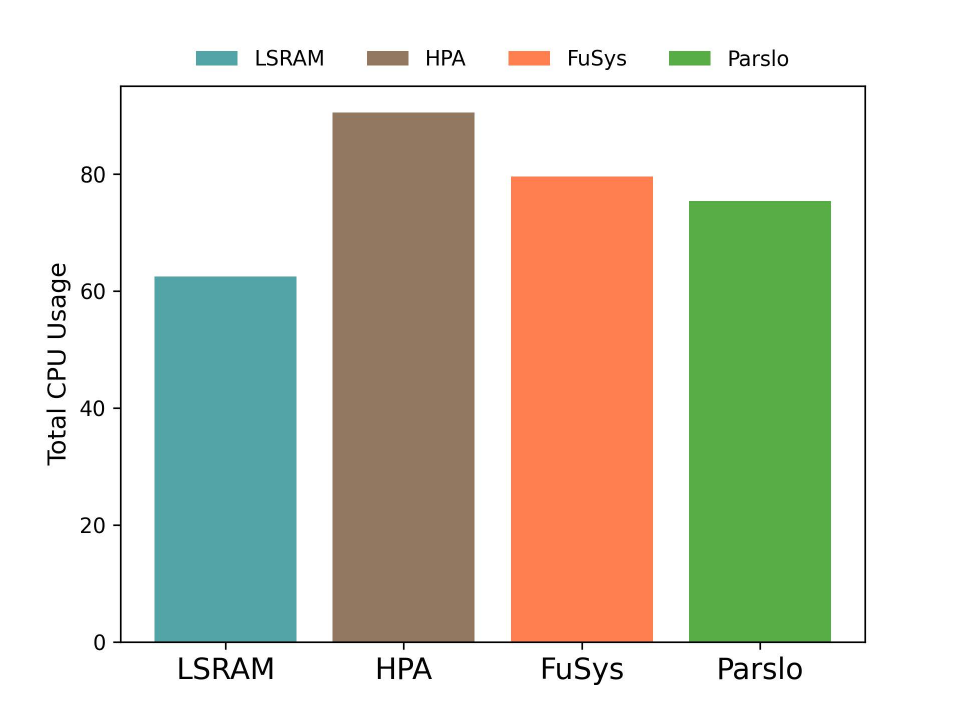}
  \caption{Total CPU.}
  \label{fig:total_cpu}
\end{subfigure}
\caption{Analysis of results}
\label{fig:result2}
\end{figure}


\textcolor{black}{To further demonstrate LSRAM's ability to maintain stable response times during runtime. With numerical analysis for RT, we can note that the Cumulative Distribution Function (CDF) graph depicted in Figure.~\ref{fig:result2}a shows that LSRAM's 20\% RT are around 190ms, and 80\% of the RT are less than 250ms, with the maximum value of 260ms. Although 80\% of the RT of the other three methods are around 200ms, all of them have exceptionally maximum RT (tail latency), e.g. the tail latency of FuSys is around 400ms. LSRAM maintains the RT within the acceptable SLO, which ensures the QoS and saves resources compared with other three baselines. The main reason is that LSRAM constantly adapts to the current environment and adjusts the resource allocation threshold through the prediction model and the SLO resource update model.}

We also further analyze the total resources consumed of  each baseline, LSRAM has saved the total resource consumption by 30\% compared with HPA, 21.4\% compared with FuSys, and 17\% compared with Parslo, as shown in Figure.~\ref{fig:result2}b. LSRAM's lightweight SLO allocation model and SLO resource update algorithm enable the framework to compute SLO resource schemes that match the current cluster environment in real-time without significant resource consumption. The SLO resource update algorithm dynamically adjusts the allocation scheme based on the historical load distribution and microservice characteristics. This ensures that the overall SLO resource scheme continuously remains aligned with the current cluster environment, which is the main reason that LSRAM is capable of maintaining low resource consumption.
\begin{figure}[htbp]
    \centering
    \begin{subfigure}[b]{0.35\linewidth}
        \includegraphics[width=\linewidth]{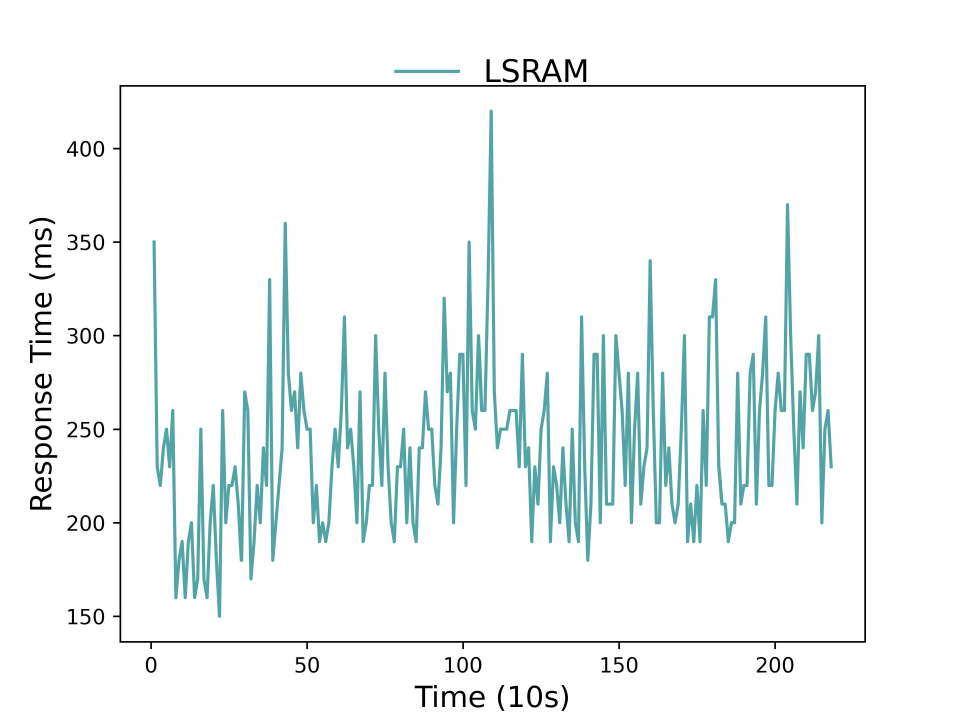}
        \caption{LSRAM}
        \label{fig:1}
    \end{subfigure}
    \begin{subfigure}[b]{0.35\linewidth}
        \includegraphics[width=\linewidth]{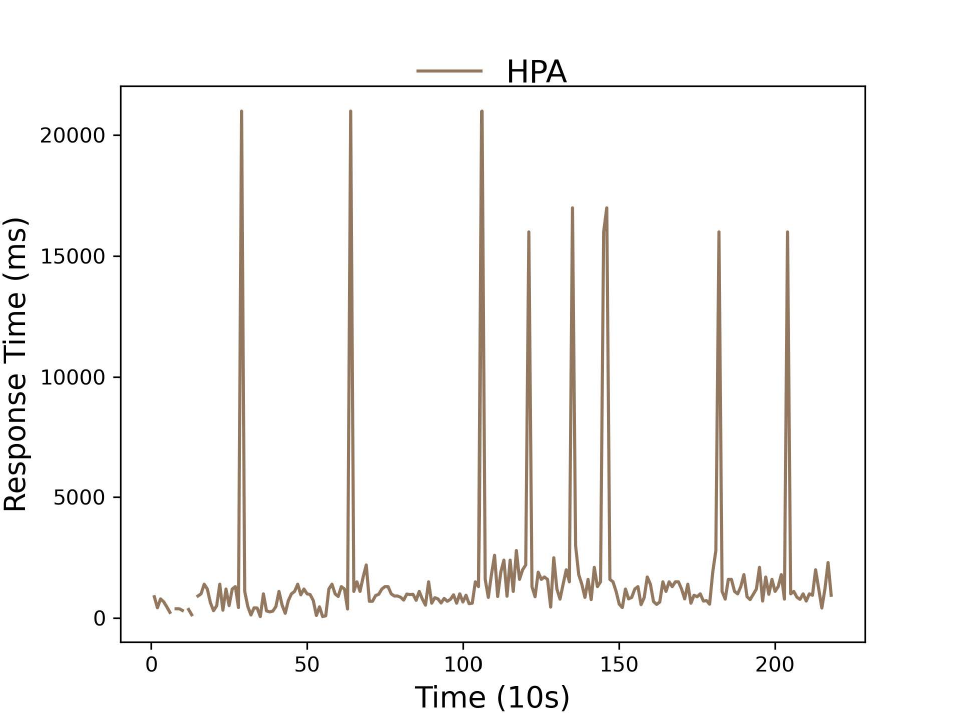}
        \caption{HPA}
        \label{fig:2}
    \end{subfigure}
    \begin{subfigure}[b]{0.35\linewidth}
        \includegraphics[width=\linewidth]{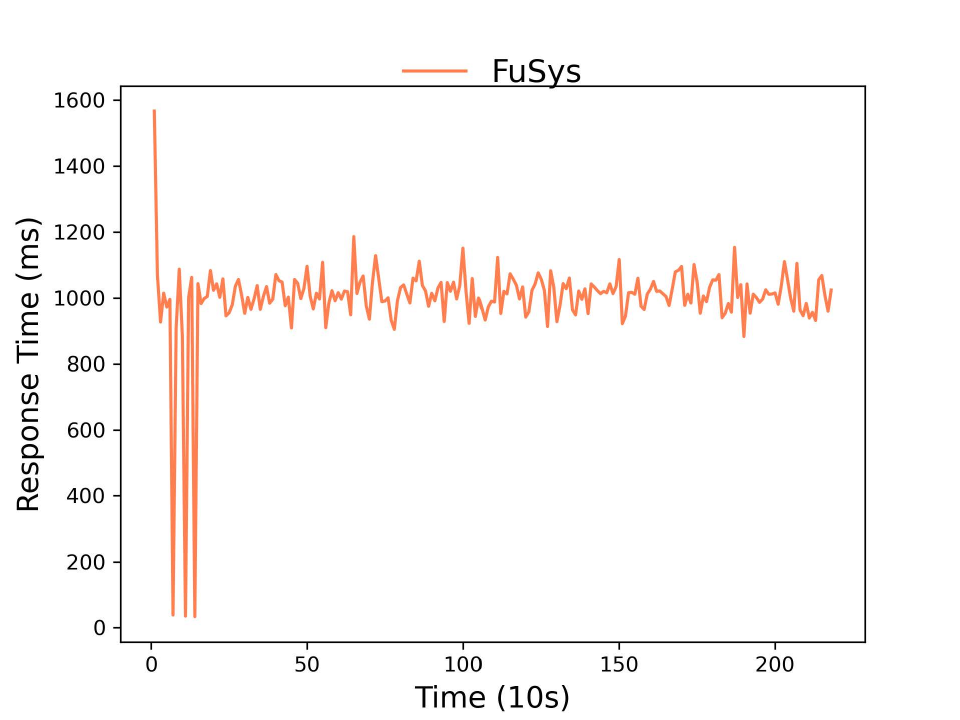}
        \caption{FuSys}
        \label{fig:3}
    \end{subfigure}
    \begin{subfigure}[b]{0.35\linewidth}
        \includegraphics[width=\linewidth]{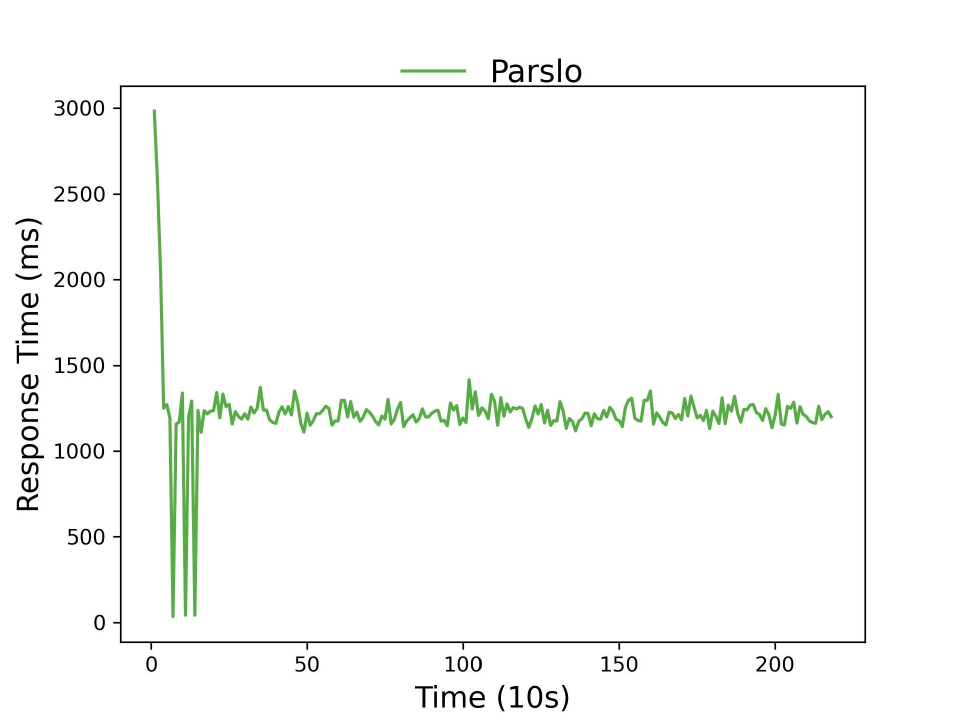}
        \caption{Parslo}
        \label{fig:4}
    \end{subfigure}
    \caption{Performance under Highly Volatile Load.}
    \label{fig:all_4}
\end{figure}

To measure LSRAM's performance under highly volatile load scenarios, we selected a steep load data pattern, as illustrated in Figure.~\ref{fig:llp4} with maximum difference between peak and bottom can be 700 requests per second (RPS). The results demonstrate that LSRAM effectively maintains users' QoS, which keeps the RT to be stable at 250ms, with maximum RT as 400ms under extreme high loads. Parslo and FuSys are more sensitive to violate loads, resulting in a congested requests processing. HPA, due to its heuristic scaling and fixed threshold, experiences severe SLO violations when there are spikes. 

In contrast, the other baselines, due to their heuristic nature, struggle to allocate resources promptly to handle such load scenarios, resulting in significant increase in RT, as depicted in Figure.~\ref{fig:all_4}. During runtime, LSRAM's load predictor accurately forecast load changes in future time period as we have trained our load prediction model to fit with bursts. However, the other baselines respond heuristically to current load changes, often failing to provide timely responses to large fluctuation ranges in loads, even leading to instances crashed. In addition to load prediction, LSRAM incorporates a mode converter that monitors load changes over time. When detecting load oscillations, LSRAM adopts a scaling mechanism to maintain SLO. This proactive approach further enhances LSRAM's ability to handle highly volatile loads effectively.

\begin{figure}[htbp!]
\centering
\begin{subfigure}[b]{0.4\linewidth}
  \centering
  \includegraphics[width=\linewidth]{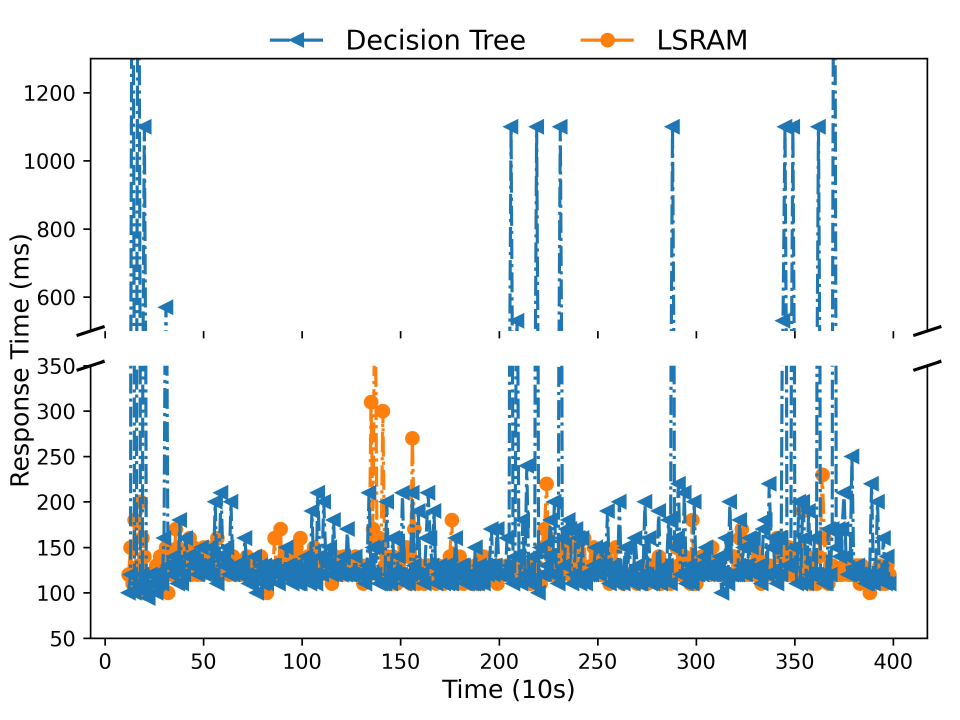}
  \caption{\textcolor{black}{Response Time.}}
  \label{fig:trans_rt}
\end{subfigure}
\begin{subfigure}[b]{0.4\linewidth}
  \centering
  \includegraphics[width=\linewidth]{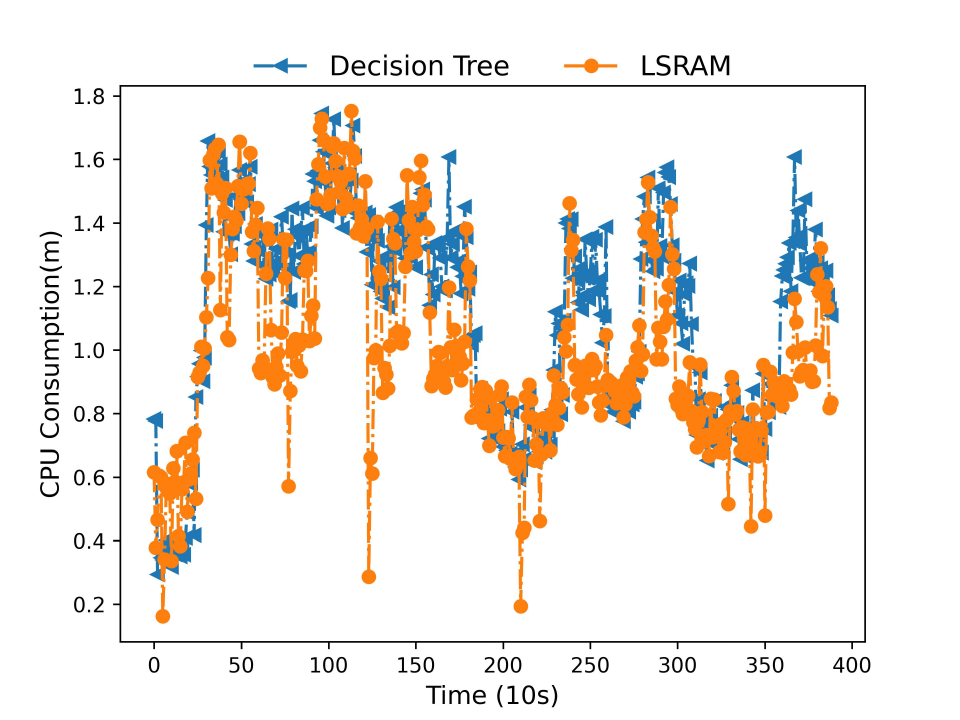}
  \caption{\textcolor{black}{CPU Resource Consumption.}}
  \label{fig:trans_cpu}
\end{subfigure}
\caption{\textcolor{black}{Comparison between different mode converter.}}
\label{fig:trans}
\end{figure}

\textcolor{black}{To validate the effectiveness of our mode converter, we compared it with current mainstream mode converter. We compared our model converter with a decision tree-based pattern converter \cite{hou2020ant}. The decision tree's depth is primarily determined by variance features, load variation trends, and the gap between maximum and minimum values. 
As shown in the experimental results in Figure.~\ref{fig:trans}, LSRAM outperforms the decision tree-based pattern converter in ensuring SLO compliance, not only providing better QoS but also conserving more system resources. During testing, the decision tree-based converter occasionally misjudged or delayed decisions, causing it to fail to switch modes promptly during certain fluctuating traffic periods, resulting in a spike in service request response times. When comparing CPU consumption, as shown in Figure.~\ref{fig:trans_cpu}, the decision tree converter also exhibited misjudgment during operation, treating non-significant load fluctuations as significant ones and triggering mode switching, leading to increased CPU resource consumption during certain periods.}

\section{Conclusions and Future Work}\label{conclu}

In this paper, we identify limitations in the current SLO resource allocation-based autoscaling framework regarding the accuracy of allocation time duration calculation and managing unexpected traffic flow. Consequently, we propose a lightweight SLO resource allocation model to swiftly compute the allocation scheme. Additionally, we introduce an update model enabling LSRAM to adjust microservices' SLO resources based on real-time cluster statuses. To handle bursty traffic and fluctuating loads effectively, we enhance the esDNN prediction model and introduce a pattern converter. Our experimental evaluation in a clustered environment demonstrates that LSRAM not only ensures effective QoS but also reduces resource usage by 17\% compared to the state-of-the-art baseline approach. 

In future work, we plan to explore the relationship between end-to-end latency and partial tail latency using deep learning methods. Specifically, we intend to explore more complex and extensive microservices architectures, where multiple tasks are dynamically scheduled and managed. This includes scenarios involving heterogeneous workloads and mixed workload patterns. Furthermore, we hope to investigate a reinforcement learning-based SLO resource allocation architecture that can quickly adapt to load and environmental changes, thereby improving system resource utilization efficiency.

\section*{AUTHOR CONTRIBUTIONS}
All authors contributed to the final manuscript by discussing the ideas and analyzing the results. Minxian Xu proposed the initial idea and guided the experimental design and manuscript writing. Kan Hu optimized the initial idea, designed and conducted the experiments, and wrote the main part of this work. Kejiang Ye and Chengzhong Xu provided insights on the suitable scenario and the scalability of the proposed approach, and improves the writing of this manuscript.

\section*{Acknowledgment}
This work is supported by National Key R \& D Program of China (No. 2021YFB3300200), the National Natural Science Foundation of China (No. 62072451, 62102408, 92267105), Guangdong Basic and Applied Basic Research Foundation (No. 2024A1515010251, 2023B1515130002), Guangdong Special Support Plan (No. 2021TQ06X990), Shenzhen Basic Research Program under grants JCYJ20220818101610023, and JCYJ20240809180935001, Shenzhen Industrial Application Projects of undertaking the National key R \& D Program of China (No. CJGJZD20210408091600002).

\hypersetup{citecolor=black}
\bibliography{wileyNJD-AMA.bib}

\end{document}